\documentclass[apj]{emulateapj}
\usepackage{natbib}
\usepackage{multirow}

\begin{document}

\title{Giant Molecular Clouds in the Early-Type Galaxy NGC4526}

\author{Dyas Utomo\altaffilmark{1}, Leo Blitz\altaffilmark{1}, Timothy Davis\altaffilmark{2,3}, Erik Rosolowsky\altaffilmark{4}, \\ Martin Bureau\altaffilmark{5}, Michele Cappellari\altaffilmark{5}, and Marc Sarzi\altaffilmark{3}}

\altaffiltext{1}{Department of Astronomy and Radio Astronomy Laboratory, University of California, Berkeley, CA 94720, USA; dyas@berkeley.edu.}
\altaffiltext{2}{European Southern Observatory, Karl-Schwarzschild-Strasse 2, 85748 Garching-bei-Muenchen, Germany.}
\altaffiltext{3}{Centre for Astrophysics Research, University of Hertfordshire, Hatfield, Herts AL1 9AB, UK.}
\altaffiltext{4}{Department of Physics, University of Alberta, 4-181 CCIS, Edmonton, AB T6G 2E1, Canada.}
\altaffiltext{5}{Sub-department of Astrophysics, Department of Physics, University of Oxford, Denys Wilkinson Building, Keble Road, Oxford OX1 3RH, UK.}

\begin{abstract}
We present a high spatial resolution ($\approx 20$ pc) of $^{12}$CO($2-1$) observations of the lenticular galaxy NGC4526. We identify 103 resolved Giant Molecular Clouds (GMCs) and measure their properties: size $R$, velocity dispersion $\sigma_v$, and luminosity $L$. This is the first GMC catalog of an early-type galaxy. We find that the GMC population in NGC4526 is gravitationally bound, with a virial parameter $\alpha \sim 1$. The mass distribution, $dN/dM \propto M^{-2.39 \pm 0.03}$, is steeper than that for GMCs in the inner Milky Way, but comparable to that found in some late-type galaxies. We find no size-linewidth correlation for the NGC4526 clouds, in contradiction to the expectation from Larson's relation. In general, the GMCs in NGC4526 are more luminous, denser, and have a higher velocity dispersion than equal size GMCs in the Milky Way and other galaxies in the Local Group. These may be due to higher interstellar radiation field than in the Milky Way disk and weaker external pressure than in the Galactic center. In addition, a kinematic measurement of cloud rotation shows that the rotation is driven by the galactic shear. For the vast majority of the clouds, the rotational energy is less than the turbulent and gravitational energy, while the four innermost clouds are unbound and will likely be torn apart by the strong shear at the galactic center. We combine our data with the archival data of other galaxies to show that the surface density $\Sigma$ of GMCs is not approximately constant as previously believed, but varies by $\sim 3$ orders of magnitude. We also show that the size and velocity dispersion of GMC population across galaxies are related to the surface density, as expected from the gravitational and pressure equilibrium, i.e. $\sigma_v R^{-1/2} \propto \Sigma^{1/2}$.
\end{abstract}


\section{Introduction}

Giant molecular clouds (GMCs) are the sites of star formation in galaxies. The existing correlation between molecular gas surface density and star formation rate \citep{wong02,bigiel08,leroy13} implies that the formation and evolution of GMCs are essential to understand the build-up of stellar masses in galaxies. However, up-to-date studies of extragalactic GMC populations are limited to Local Group galaxies; LMC \citep{fukui08,wong11}, SMC \citep{mizuno01}, M31 \citep{eros07}, M33 \citep{engargiola03,eros07b,gratier12}, and IC10 \citep{leroy06}, the nearby spirals; M64 \citep{eros05} and M51 \citep{colombo14}, and the nearby starburst; M82 \citep{keto05} and NGC253 \citep{leroy14}, due to the limited angular resolution and sensitivity of radio telescopes. Galaxies in the Local Group are mostly dwarfs with few spirals. Therefore, additional study of GMCs in the early-type galaxies, such as NGC4526, is needed to provide a comprehensive analysis of GMC properties across different galaxy environments.

There are three resolved GMC properties that we can directly measure: size, linewidth, and luminosity. The relationships between these properties were first studied by \citet{larson81}, who suggested the importance of turbulence in the stability of GMCs against self-gravity. These relations were then refined by \citet[hereafter S87]{solomon87} for GMCs in the Milky Way disk. Basically, GMC properties in the Milky Way can be described by three Larson's `laws`: (1) GMCs are gravitationally bound objects, (2) the size and velocity dispersion of GMCs follows a $\sigma_v \propto R^{0.5}$ relation, and by implication (3) the surface density of GMCs is approximately constant ($\Sigma_{\rm GMC} \approx 170 \ M_{\odot} \ {\rm pc}^{-2}$, S87). Interestingly, these relations also hold true, albeit with scatter, for extragalactic GMCs in Local Group galaxies \citep{blitz07,bolatto08,fukui10}. At face value, this suggests that GMC properties are universal.

However, further studies reveal that GMC properties can deviate from Larson's relations. \citet{heyer01} found that low mass GMCs ($M \leq 10^3 M_{\odot}$) in the outer part of the Milky Way are not gravitationally bound. Their luminous masses, inferred from the CO-to-H$_2$ conversion factor, are smaller than their virial masses. The required external pressure to bind these clouds is $P_{\rm ext}/k \sim 1 \times 10^4$ K cm$^{-3}$. Furthermore, when re-examining S87 clouds using more sensitive instruments, \citet{heyer09} found that the surface density of Milky Way disk GMCs varies from $\sim 10$ to $200 \ M_{\odot} \ {\rm pc}^{-2}$, and they deviate from gravitational equilibrium. \citet{field11} suggest that these clouds may be in pressure virial equilibrium, where the clouds' mass and radius are described by a Bonnor-Ebert sphere with various external pressures \citep{bonnor56,ebert55}. In addition, several authors \citep[e.g.][]{kegel89,ballesteros02} argue that the observed constancy of surface density might be affected by observational biases.

If this is really the case, then what factors set the different properties of GMCs? These parameters may be external (environmental) effects, such as hydrostatic pressure \citep{elmegreen93,blitz04,meidt13}, interstellar radiation field strength \citep{mckee89}, and shear from galaxy rotation \citep{koda09,miyamoto14}, or internal, such as feedback of the star formation that is embedded inside GMCs \citep{mckee89}. To answer this question, we need a complete sample of GMCs across different environments: from bulge to spiral-arm and inter-arm regions, from late-type to early-type galaxies, and from low to high metallicity galaxies.

In this respect, we analyze the GMC properties in the bulge of NGC4526, an S0-type galaxy in the Virgo cluster. NGC4526 is unusual because all of the CO in the galaxy has been observed at a linear resolution of $\approx 20$ pc, sufficient to resolve Milky Way sized GMCs. The galaxy has prominent central dust lanes with mass $\sim 10^7 M_{\odot}$ \citep{ciesla14} and supersolar metallicity \citep[log$(Z/Z_{\odot}) \approx 0.2$;][]{davis13c}, but lack of star formation \citep[SFR $\approx 0.03 \ M_{\odot} \ {\rm yr}^{-1}$;][]{amblard14} and devoid of atomic gas \citep[$M_{\rm HI} < 1.9 \times 10^7 M_{\odot}$;][]{lucero13}. The HI deficiency in this galaxy may be caused by ram pressure or evaporation by hot gas (as the galaxy resides in the Virgo cluster), or by abrupt conversion of HI into molecular gas due to high pressure \citep{elmegreen93}. There is no indication of recent tidal interaction \citep{young08}, suggesting that this mechanism is not the primary cause of HI deficiency in NGC4526.

In fact, the molecular gas in NGC4526 is confined within the central $\sim 1$ kpc region \citep[the top panel of Figure 1;][]{davis13b}. The central regions of galaxies, such as in the Milky Way, tend to have high interstellar pressures \citep[$\sim 5 \times 10^6 \ {\rm K \ cm}^{-3}$;][]{spergel92}, strong magnetic fields \citep[$\sim 1$ mG;][]{zadeh87}, and lower than expected star formation rate \citep{longmore13}. These properties offer a unique environment for GMCs in NGC4526, significantly different than those studied in other galaxies.

This paper is organized as follows. In $\S 2$ we describe the data and methodology to identify GMCs in NGC4526. The properties and kinematics of the GMCs are reported in $\S 3$ and $\S 4$, respectively, and catalogued in Table 1. We discuss the pressure balance of GMCs in $\S 5.1$, the Larson's `laws` in $\S 5.2$ and $\S 5.3$, and the environmental effects to the GMC properties in $\S 5.4$. Lastly, we summarize our findings in $\S 6$.

\section{Data and Methodology}

\subsection{Data Descriptions}

\begin{figure*}
%\figurenum{text}
\epsscale{1.16}
\plotone{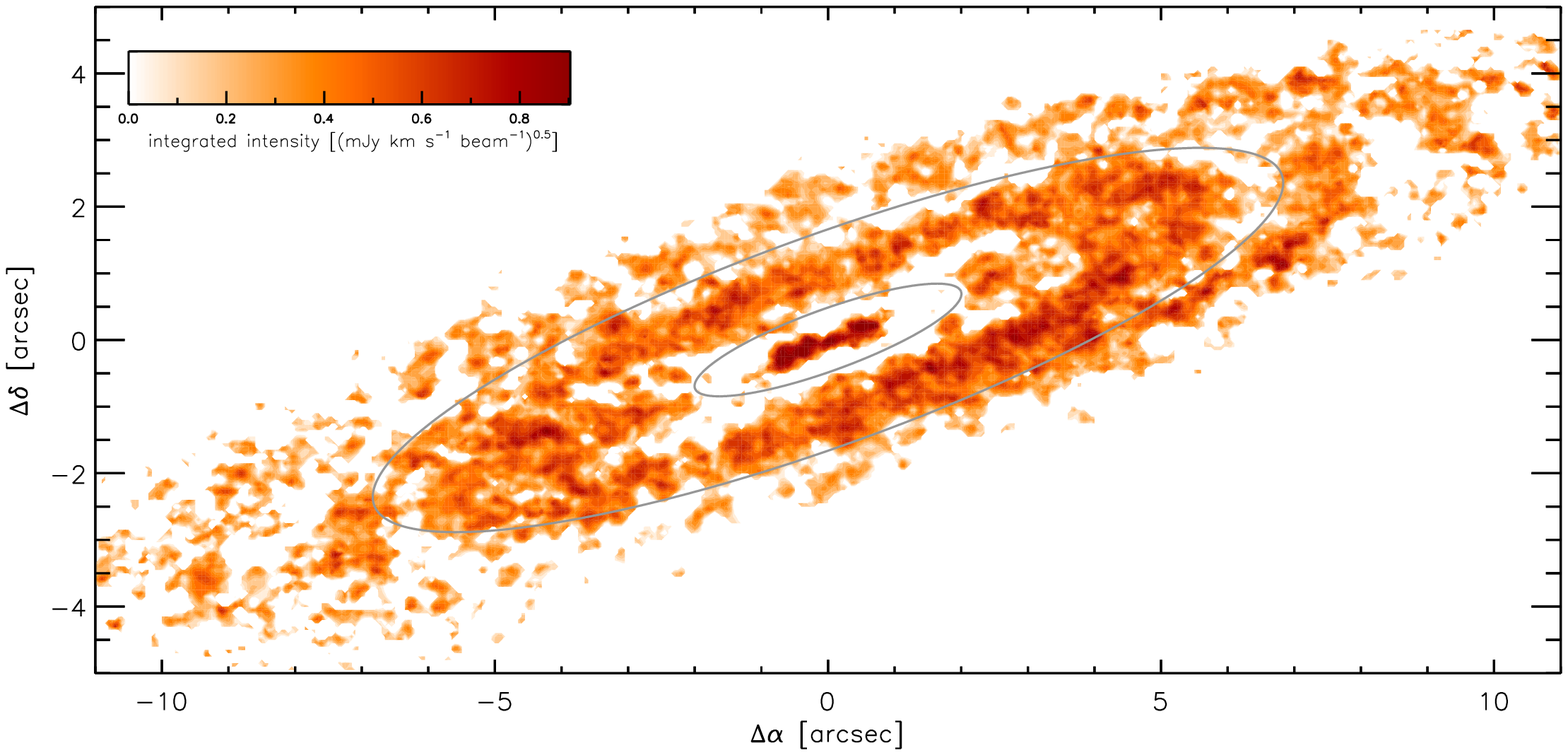}
\epsscale{1.18}
\plotone{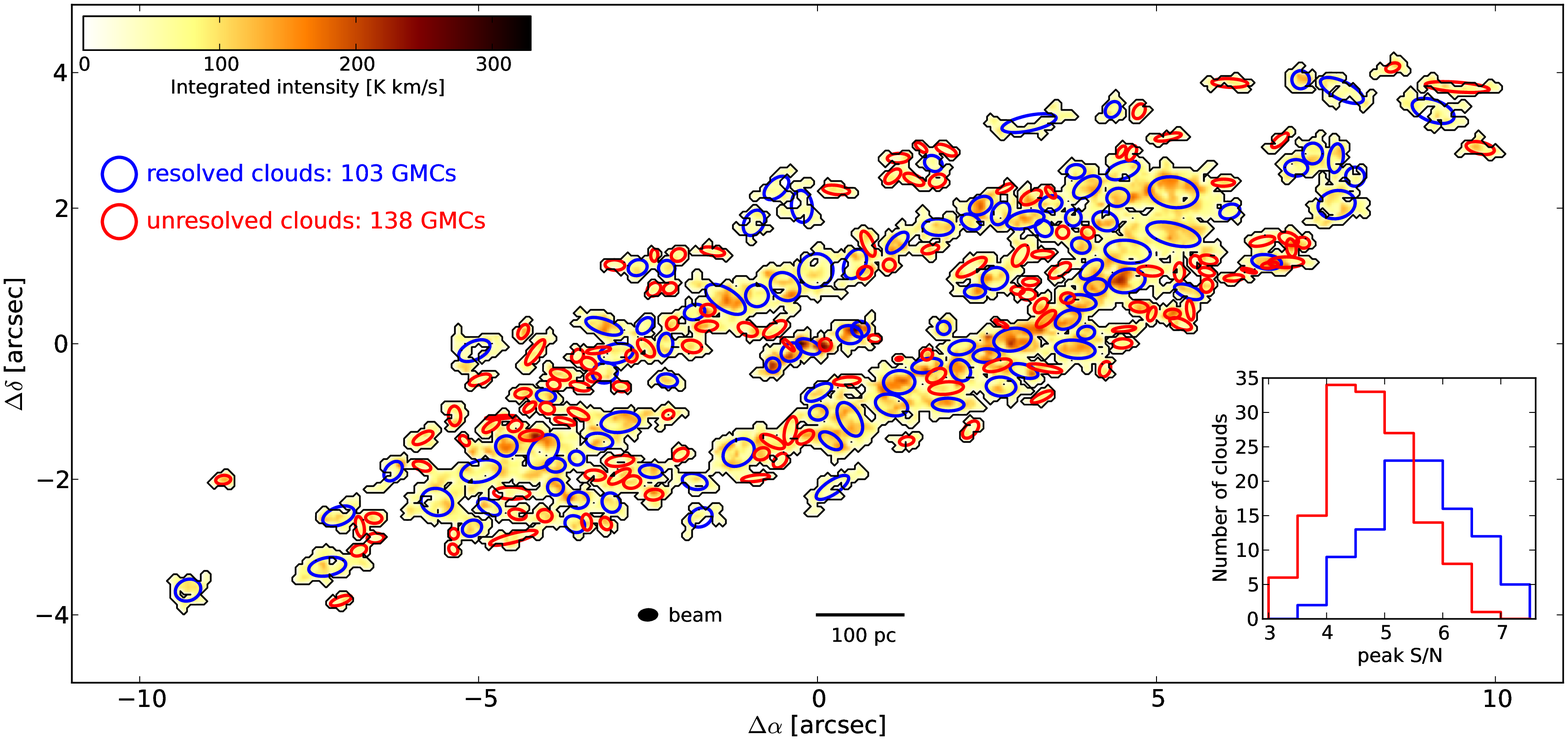}
\label{fig:1}
%\plottwo{epsfile}{epsfile}
\caption{Top: The integrated intensity map of NGC4526. The map is created by applying a Gaussian fit to each spectrum in the data cube. We exclude any Gaussian that has peak less than $2.5 \sigma_{\rm rms}$. The ellipses divide the CO emission into three zones: inner region, molecular ring, and outer region. Bottom: Identified GMCs in NGC4526 overploted on the masked integrated intensity map. The mask covers regions with connected emission above $2\sigma_{\rm rms}$ and having at least one pixel with $3\sigma_{\rm rms}$. The beam size and the projected physical size are indicated. The blue and red circles mark the location of the resolved and unresolved clouds, respectively. The distribution of the peak S/N of clouds is shown as an inset in the bottom right corner.}
\end{figure*}

NGC4526 was observed in the $^{12}$CO($2-1$) line (230 GHz or 1.3 mm) using the Combined Array for Research in Millimeter-wave Astronomy (CARMA) in A, B, and C configurations \citep{bock06}. The data were taken as part of the mm-Wave Interferometric Survey of Dark Object Masses (WISDOM) project. Results for the innermost CO were reported by \citet{davis13b}, who showed that the kinematics of the central CO imply the presence of a $4.5 \times 10^8 M_{\odot}$ supermassive black hole.

The beam width of the observations is $0.278 \times 0.173$ arcsec$^2$ and the spectral resolution (after Hanning smoothing) is 10 km s$^{-1}$. This beam width covers $5.56 \times 3.46$ pixels and corresponds to a projected physical size of $\approx 22 \times 14$ pc$^2$ at the adopted distance of 16.4 Mpc \citep{tonry01}. These high resolution data enable us to resolve individual GMCs and measure their properties, since the typical Milky Way's GMC sizes are $\sim 50$ pc \citep[e.g.][]{blitz93}.

The noise in our data is not uniform, with higher noise appearing at the corners of the data cube. The overall pixel-by-pixel root-mean-square (rms) noise $\sigma_{\rm rms}$ distribution is a positively-skewed Gaussian with minimum, average, and maximum values of 0.33, 0.71, and 1.33 K, respectively.

\subsection{Methodology}

We identify GMC candidates in NGC4526 using the modified CLUMPFIND algorithm \citep{williams94}, implemented in the CPROPS program \citep[hereafter RL06]{eros06}. The main goal of this program is to identify all real clouds and minimize false detections due to noise fluctuations. Descriptions of the CPROPS program, together with our chosen values of the input parameters of the program, are given in Appendix A and B.

As a result of the CPROPS analysis, 241 GMCs are identified in NGC4526, of which 103 are resolved. We assume all GMCs are real since the probability of false detections is very small (Appendix C). In the bottom panel of Figure 1, we show the integrated CO emission of connected regions that have brightness temperatures $T_{\rm b} > 2\sigma_{\rm rms}$ and have at least one pixel with $T_{\rm b} \geq 3\sigma_{\rm rms}$. The locations of resolved and unresolved clouds are marked as blue and red ellipses, respectively. The peak S/N distribution of identified GMCs is shown as an inset. The mean peak S/N of resolved and unresolved clouds is 5.6 and 4.7, respectively.

Most GMCs are located in the molecular ring, a few hundred parsecs from the galactic center (Figure 1). This molecular ring is the largest contiguous CO emission in our data. In addition, there are a few clouds located in the central region. The outer region of the molecular gas exhibits a spiral-arm structure, possibly with an outer pseudo-ring, which fragments into smaller structures consisting of one or multiple GMCs. Except for one cloud that is described below, all identified GMCs are within 900 pc of the galactic center, i.e. inside the bulge of NGC4526. Note that our primary beam covers all of the CO emission in the galaxy, so our GMC catalog is complete.

There is one unresolved cloud which is located on the edge of the data cube (not shown in Figure 1), out of the plane of the CO emission. We overplotted the location of this cloud with the HST archival image of the galaxy. Although this cloud is likely to be real (Appendix C), its distance is uncertain, i.e. it may be located outside the galaxy. We exclude this cloud from the following analysis but keep it in the catalog (as cloud no. 80 in Table 1). Inclusion of this cloud does not alter the conclusions of our analysis.

\section{Cloud Properties}

\subsection{Definition of GMC Properties}

Cloud properties, such as position, size, velocity dispersion, luminosity, and mass, are catalogued in Table 1. Here, we briefly describe the method used to measure the cloud properties. Full explanations of the method are given in RL06.

\begin{table*}
\caption{NGC4526 Cloud Properties}
\centering
\begin{tabular}{ c c c c c c c c c c c c c c c c }
\tableline
ID & RA(2000) & Dec(2000) & $V_{\rm LSR}$\tablenotemark{a} & $R$\tablenotemark{b} & $\delta R$\tablenotemark{b} & $\sigma_v$\tablenotemark{a} & $\delta \sigma_v$\tablenotemark{a} & $L$\tablenotemark{c} & $\delta L$\tablenotemark{c} & $M_{\rm lum}$\tablenotemark{d} & $\delta M_{\rm lum}$\tablenotemark{d} & S/N & $T_{b, {\rm max}}$ & $\Omega_{\rm shear}$\tablenotemark{e} & $d$\tablenotemark{f} \\[0.5ex]
& [h:m:s] & [$^{\circ}:':''$] &  & [pc] & [pc] &  &  & & & & & & [K] & & [pc]  \\[0.5ex]
\tableline
1 & 12:34:3.5 & 7:41:54.7 & 276.0 & 24.00 & \phantom{1}7.65 & 6.90 & 0.48 & 3.76 & 1.24 & 1.88 & 0.62 & 5.0 & \phantom{1}7.7 & 0.52 & 667 \\[0.5ex]
2 & 12:34:3.5 & 7:41:54.9 & 283.4 & \dots &\dots & 6.14 & 1.30 & 0.67 & 0.60 & 0.33 & 0.30 & 4.1 & \phantom{1}6.3 & 0.55 & 619 \\[0.5ex]
3 & 12:34:3.4 & 7:41:55.1 & 286.7 & \dots & \dots & 6.37 & 1.21 & 0.27 & 0.19 & 0.14 & 0.09 & 3.4 & \phantom{1}4.7 & 0.57 & 587 \\[0.5ex]
4 & 12:34:3.6 & 7:41:54.3 & 279.6 & 22.16 & \phantom{1}5.54 & 5.42 & 0.55 & 2.35 & 0.70 & 1.18 & 0.35 & 6.6 & 10.4 & 0.44 & 798 \\[0.5ex]
5 & 12:34:3.5 & 7:41:55.2 & 286.7 & \dots & \dots & 5.64 & 1.55 & 0.71 & 0.68 & 0.35 & 0.34 & 4.6 & \phantom{1}6.7 & 0.57 & 583 \\[0.5ex]
6 & 12:34:3.4 & 7:41:55.4 & 293.2 & \dots & \dots & 9.74 & 2.35 & 1.13 & 0.60 & 0.57 & 0.30 & 5.2 & \phantom{1}6.7 & 0.58 & 563 \\[0.5ex]
7 & 12:34:3.5 & 7:41:55.4 & 298.3 & 22.01 & \phantom{1}5.52 & 7.23 & 0.96 & 2.39 & 0.76 & 1.20 & 0.38 & 5.7 & \phantom{1}8.1 & 0.56 & 599 \\[0.5ex]
8 & 12:34:3.4 & 7:41:55.1 & 298.5 & \dots & \dots & 6.28 & 1.67 & 0.67 & 0.99 & 0.33 & 0.50 & 3.6 & \phantom{1}5.1 & 0.58 & 559 \\[0.5ex]
9 & 12:34:3.4 & 7:41:55.6 & 297.2 & 29.60 & \phantom{1}6.81 & 7.25 & 0.63 & 5.75 & 1.35 & 2.88 & 0.68 & 6.2 & \phantom{1}9.0 & 0.61 & 492 \\[0.5ex]
10 & 12:34:3.3 & 7:41:55.5 & 311.0 & 12.11 & \phantom{1}9.12 & 8.01 & 1.41 & 1.42 & 1.35 & 0.71 & 0.68 & 4.4 & \phantom{1}6.4 & 0.61 & 480 \\[0.5ex]
11 & 12:34:3.3 & 7:41:55.7 & 314.9 & \dots & \dots & 6.27 & 0.55 & 1.05 & 0.68 & 0.53 & 0.34 & 5.1 & \phantom{1}7.1 & 0.62 & 440 \\[0.5ex]
12 & 12:34:3.3 & 7:41:56.1 & 309.0 & 28.58 & \phantom{1}5.57 & 7.12 & 0.33 & 5.52 & 1.16 & 2.77 & 0.58 & 7.3 & 10.2 & 0.63 & 424 \\[0.5ex]
13 & 12:34:3.3 & 7:41:56.4 & 323.3 & 31.38 & \phantom{1}5.24 & 7.81 & 0.86 & 5.84 & 1.08 & 2.93 & 0.54 & 7.3 & 10.3 & 0.66 & 348 \\[0.5ex]
14 & 12:34:3.3 & 7:41:55.2 & 323.5 & 13.38 & \phantom{1}8.55 & 8.21 & 4.00 & 2.01 & 0.87 & 1.01 & 0.44 & 5.1 & \phantom{1}7.4 & 0.59 & 543 \\[0.5ex]
15 & 12:34:3.3 & 7:41:56.4 & 333.7 & 18.72 & \phantom{1}5.54 & 9.56 & 2.32 & 3.18 & 0.87 & 1.59 & 0.44 & 6.6 & \phantom{1}8.9 & 0.64 & 388 \\[0.5ex]
16 & 12:34:3.2 & 7:41:56.5 & 344.0 & 16.65 & \phantom{1}7.01 & 8.75 & 2.44 & 2.00 & 0.83 & 1.00 & 0.41 & 5.8 & \phantom{1}7.8 & 0.76 & 293 \\[0.5ex]
17 & 12:34:3.2 & 7:41:56.8 & 328.9 & \dots & \dots & 7.64 & 1.41 & 0.29 & 0.21 & 0.14 & 0.11 & 4.1 & \phantom{1}5.8 & 0.70 & 320 \\[0.5ex]
18 & 12:34:3.4 & 7:41:56.1 & 335.7 & \dots & \dots & 7.39 & 3.21 & 0.80 & 0.91 & 0.40 & 0.46 & 4.4 & \phantom{1}6.0 & 0.60 & 499 \\[0.5ex]
19 & 12:34:3.3 & 7:41:56.2 & 330.3 & 10.48 & \phantom{1}7.41 & 8.53 & 1.50 & 1.47 & 0.64 & 0.74 & 0.32 & 5.5 & \phantom{1}7.9 & 0.66 & 360 \\[0.5ex]
20 & 12:34:3.2 & 7:41:56.3 & 337.5 & \phantom{1}7.57 & 11.12 & 6.44 & 1.74 & 0.96 & 0.97 & 0.48 & 0.49 & 3.9 & \phantom{1}5.9 & 0.68 & 336 \\[0.5ex]
\tableline
\end{tabular}
\tablenotetext{1}{Units are km s$^{-1}$.}
\tablenotetext{2}{The size of unresolved clouds is less than the linear size of the beam, denoted as $R = \dots$.}
\tablenotetext{3}{Units are $10^5$ K km s$^{-1}$ pc$^2$.}
\tablenotetext{4}{Units are $10^6 M_{\odot}$.}
\tablenotetext{5}{Units are km s$^{-1}$ pc$^{-1}$.}
\tablenotetext{6}{Distance from the center of NGC4526, assuming clouds are in the plane of the galaxy with axis ratio of 0.216 and position angle of $20.2^{\circ}$.}
\tablecomments{Table 1 is published in its entirety in the electronic edition of the Astrophysical Journal. A portion is shown here for guidance regarding its form and content.}
\end{table*}

The cloud size, $R$, is measured using the {\it deconvolved} second-moment:
\begin{equation}
R = \eta \sqrt{(\sigma_{\rm maj}^2 [0 \ {\rm K}] - \sigma_{\rm beam}^2)^{1/2} (\sigma_{\rm min}^2 [0 \ {\rm K}] - \sigma_{\rm beam}^2)^{1/2}},
\end{equation}
where $\eta$ is a factor that depends on the density distribution of spherically symmetric clouds. A uniform sphere has $\eta = \sqrt{5}$, while an isothermal sphere has $\eta = 3$. Here, we adopt $\eta = 1.19$, a value from S87, to make it consistent with previous studies. The major, $\sigma_{\rm maj}$, and minor, $\sigma_{\rm min}$, dispersions are the spatial second moments, weighted by the intensity, along the major and minor axis of the clouds, respectively. Both $\sigma_{\rm maj}$ and $\sigma_{\rm min}$ are extrapolated to zero intensity ($0 \ {\rm K}$) to avoid bias due to the limited sensitivity of the instrument. Deconvolution is applied, by the inclusion of the $\sigma_{\rm beam}$ terms in equation (1), to avoid bias due to the finite beam resolution (RL06). Since the beam is not circular, we take $\sigma_{\rm beam}$ as the geometrical mean of the major and minor axes of the beam. The uncertainty, $\delta R$, is determined using bootstrap resampling.

In order to measure the clouds' velocity dispersions, we attempt Gaussian fitting to the composite spectrum of each cloud through the following steps. First, we calculate the offset of the mean velocity at all lines of sight within the cloud $(x_i,y_i)$, with respect to the mean velocity at the center of the cloud $(x_0,y_0)$. This offset is caused by large scale motions, such as the cloud's rotation or shear due to galactic rotation.  Then, we shift each line of sight velocity spectrum to match the mean velocity of the central position of the cloud. Except for the innermost clouds, this shift removes any velocity dispersions due to large scale motions, and leaves only turbulence and thermal broadening as sources of velocity dispersions. To make a composite spectrum, we take the average velocity profile from each line of sight. Finally, we fit the composite spectrum with a Gaussian. The standard deviation of the Gaussian fit is taken as the velocity dispersion, $\sigma_v$, of the cloud. We take the uncertainty, $\delta \sigma_v$, from the bootstrap resampling.

RL06 shows that measurements of velocity dispersion suffer bias towards higher values due to the finite spectral resolution of the instruments. Therefore, we take the deconvolved value, $\sigma_{v,{\rm dc}}$, to remove this bias using the same prescription as in RL06:
\begin{equation}
\sigma_{v,{\rm dc}} = \left(\sigma_v^2 - \frac{\delta v^2}{2\pi}\right)^{1/2},
\end{equation}
where $\delta v$ is the spectral resolution, and $\delta v \ (2\pi)^{-0.5}$ is the standard deviation of a Gaussian that has an integrated area equal to a spectral channel with width $\delta v$. For brevity, we refer to $\sigma_{v,{\rm dc}}$ as $\sigma_v$. Note that the deconvolved values of $\sigma_v$ and $R$ are always smaller than their measured values (c.f. equation 1 and 2). Thus, any clouds that are barely resolved would have the deconvolved value smaller than the resolution.

The cloud luminosity is the integrated CO flux over the position-position-velocity volume occupied by the cloud (RL06):
\begin{equation}
L_{\rm CO(2-1)} =\left(\frac{\pi}{180 \times 3600}\right)^2 D^2 \sum_i T_{{\rm b},i} \ \delta x \ \delta y \ \delta v,
\end{equation}
where $T_{{\rm b},i}$ is the brightness temperature of the $i$th pixel in K, $\delta x$ and $\delta y$ are the pixel sizes in arcsec, $D$ is the distance to NGC4526 in pc, and $L_{\rm CO}$ is the cloud luminosity in K km s$^{-1}$ pc$^2$. The luminosity is extrapolated to 0 K intensity as described in RL06. The uncertainty, $\delta L$, is determined using a bootstrap resampling method. There is an additional $\sim 20\%$ (absolute, but systematic) flux calibration uncertainty that we do not include in the analysis.

The luminosity is then converted to mass using the Milky Way's CO($1-0$)-to-H$_2$ conversion factor $X_{\rm CO} = 2 \times 10^{20} \ {\rm cm}^{-2} \ ({\rm K \ km \ s}^{-1})^{-1}$, which is assumed to be constant throughout the galaxy. This is a reasonable assumption since $X_{\rm CO}$ does not vary significantly due to metallicity in the supersolar metallicity regime \citep[and references therein]{bolatto13}. The ratio of $^{12}$CO($1-0$) to $^{12}$CO($2-1$) intensity in NGC4526 is 1.15 \citep{crocker12}. We refer to this $X_{\rm CO}$-derived mass as the luminous mass: $M_{\rm lum}/M_{\odot} = (4.4 \times 1.15) \ (L_{\rm CO(2-1)}/{\rm K \ km \ s}^{-1}{\rm pc}^2)$, which takes into account the mass contribution by helium. The factor of 4.4 comes from our adopted $X_{\rm CO}$ value.

\begin{figure}
%\figurenum{text}
\epsscale{1.2}
\plotone{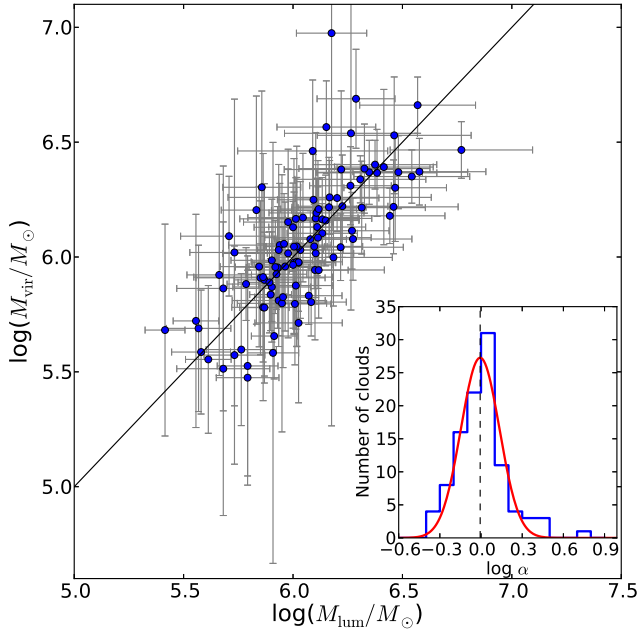}
\label{fig:2}
%\plottwo{epsfile}{epsfile}
\caption{Correlation between virial and luminous masses of GMCs. The solid line is the one-to-one relationship. The distribution of log $\alpha$ with a log-normal fit is shown as an inset. The mean of the log-normal fit is $\alpha \approx 0.99$ with a standard deviation of 0.14 dex. Thus, the GMC population in NGC4526 is in gravitational equilibrium.}
\end{figure}

We calculate each cloud's distance from the center of NGC4526 using the assumption that they are located in the plane of the galaxy with axis ratio of 0.216 and position angle of $290.2^{\circ}$. The position angle is measured from the north, counter-clockwise to the receding part of the kinematical major axis of the galaxy. The axis ratio and position angle are calculated using part of the multi-Gaussian expansion (MGE) fit of \citet{cappellari02}.

\subsection{Gravitational Equilibrium of Clouds}

The mass of a gravitationally bound cloud is given by \citep[hereafter BM92]{bertoldi92}
\begin{equation}
M_{\rm vir} = \frac{5 \sigma_v^2 R}{G} \approx 1048 \ M_{\odot} \left(\frac{\sigma_v}{\rm km \ s^{-1}}\right)^2 \left(\frac{R}{\rm pc}\right),
\end{equation}
where $\sigma_v$ is the 1-D velocity dispersion of the CO line. We refer to this mass as the virial mass. If the luminous mass is equal, or comparable, to the virial mass, then the cloud is in gravitational equilibrium, where the kinetic energy balances its self-gravity. If the luminous mass is smaller than the virial mass, then, in addition to gravity, the clouds must be held together by the external pressure of the ambient medium, $P_{\rm ext}$, to reach dynamical equilibrium. Such clouds are pressure-bound clouds.

We can also define the virial parameter as the ratio between twice the kinetic energy and the gravitational energy,
\begin{equation}
\alpha \equiv \frac{5 \sigma_v^2 R}{G M_{\rm lum}} = \frac{M_{\rm vir}}{M_{\rm lum}}.
\end{equation}
According to BM92, clouds with $\alpha \approx 1.13$ are gravitationally-bound and clouds with $\alpha \gg 1$ are pressure-bound. In Figure 2, we plot the luminous versus virial masses of the resolved clouds, together with the distribution of log $\alpha$ as an inset. A log-normal fit to the distribution yields a mean $\alpha = 0.99 \pm 0.02$ and a standard deviation of 0.14 dex. Roughly 99\% of the resolved clouds have $0.3 \leq \alpha \leq 3$, and 89\% of the resolved clouds have $0.5 \leq \alpha \leq 2$. Therefore, the GMC population in NGC4526 is in a state of gravitational equilibrium.

In equation (4), we assume that all clouds are spherically symmetric and have a uniform density distribution. If the clouds were isothermal spheres, then the virial masses would be 60\% lower than our calculation. Moreover, the uncertainty in $X_{\rm CO}$ also affects the luminous mass measurements. From observations of local galaxies, the typical uncertainty in $X_{\rm CO}$ is about 0.3 dex \citep[and references therein]{bolatto13}. Taken all together, this introduces an uncertainty of $\sim 0.5$ dex in the worst case.

Variations of the input parameters of the CPROPS program do not affect our result that the cloud population is gravitationally-bound. However, we have to keep in mind that we do not yet take into account the magnetic pressure and rotation of the clouds. The effect of rotation is discussed in $\S 4.3$. Furthermore, we find no correlation between the mass of a GMC and its distance from the galactic center, possibly because all GMCs are distributed in a small region (within a radius of 900 pc) inside the bulge of the galaxy, so environmental variations from GMC to GMC, such as ambient pressure and the interstellar radiation field, are small.

\subsection{Cloud Mass Distribution}

\begin{figure}
%\figurenum{text}
\epsscale{1.1}
\plotone{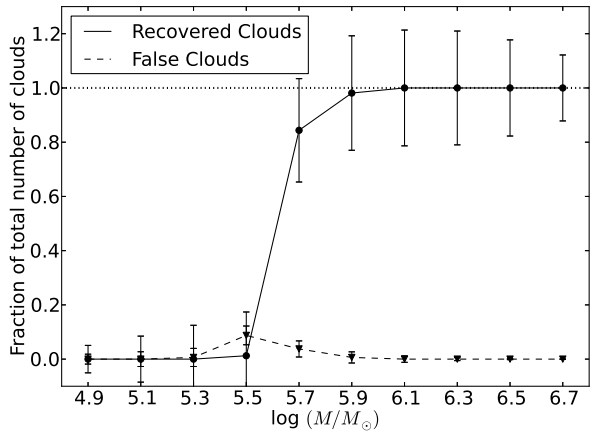}
\label{fig:2}
%\plottwo{epsfile}{epsfile}
\caption{Fraction of recovered clouds as a function of cloud mass, from the simulation described in the text. Simulated clouds with $M \geq 5 \times 10^5 M_{\odot}$ are well recovered by the program (solid line), contaminated by only a small fraction of false clouds (dashed line). The fraction of false clouds is negligible for low mass regime because these clouds are too small and faint, and hence undetected by the CPROPS program. We adopt log$(M/M_{\odot}) = 5.7$ as the completeness level of our observations.}
\end{figure}

\begin{figure*}
%\figurenum{text}
\epsscale{1.2}
\plotone{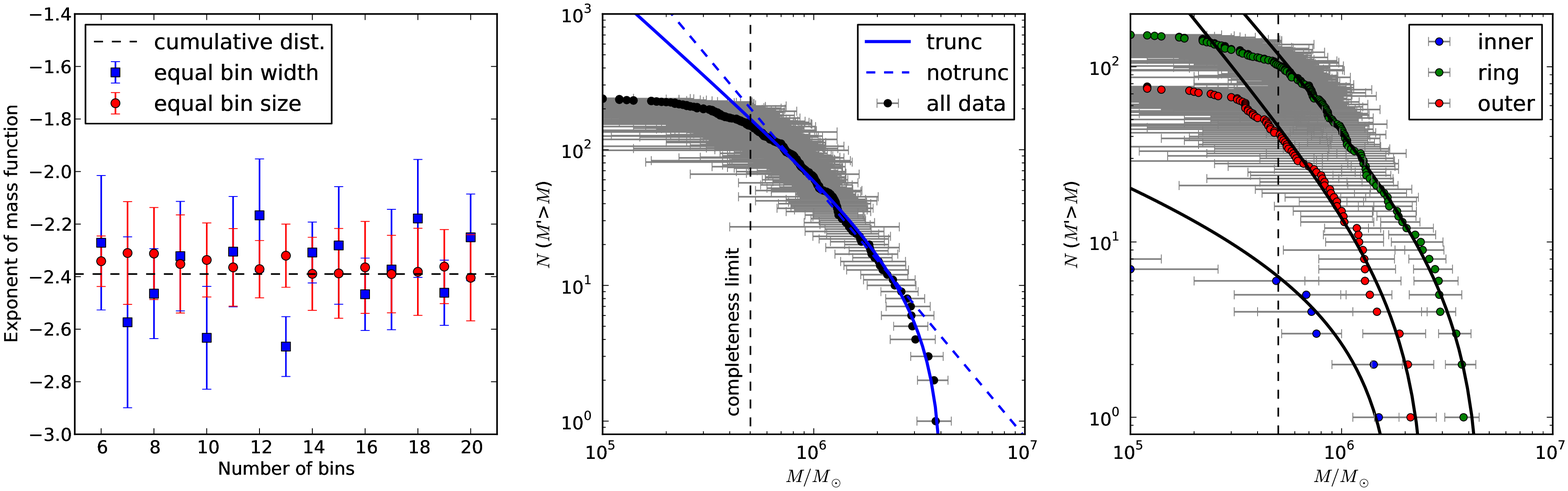}
\label{fig:4}
%\plottwo{epsfile}{epsfile}
\caption{Left: Best fit slope of the mass function against the number of bins. The equal bin-width method (blue squares) has large scatter due to the choice of the number of bins, while the equal bin-size method (red circles) is in agreement with the cumulative distribution mass function (dashed line). Middle: Fits to the cumulative mass function with and without truncation. The data favor the truncated (solid blue curve) over the non-truncated (dashed blue curve) mass function. Right: Cumulative mass distribution of the inner region (blue), molecular ring (green), and outer region (red), with the overlaid truncated fits.}
\end{figure*}

We use the luminous mass to determine the mass function because it is well-defined even for unresolved clouds. Since the GMC population in NGC4526 is in gravitational equilibrium ($M_{\rm lum} \approx M_{\rm vir}$), we should not expect variation of the mass function between the two mass measurements. We determine the mass function using three different methods: equal bin-width, equal bin-size, and the cumulative distribution function. All measurements are taken from the most massive clouds in our sample ($M_{\rm lum} \approx 5.9 \times 10^6 M_{\odot}$) down to the completeness level of the observations.

To determine the completeness level of our observations, we create simulated Gaussian clouds. Their properties are related through known scaling relationships: $\sigma_v \propto R^{0.5}$ and $M \propto R^2$ (e.g. S87). To mimic the observed data, we add the typical noise of our observations into the simulated data cubes. In total we consider 1600 mock clouds with log($M/M_{\odot}$) ranging from 4.9 to 6.7, with an increment of 0.2, and feed these mock clouds into the CPROPS program. A cloud is defined as recovered if its location in the data cube is within one beamwidth of its input location. Otherwise, this cloud is defined as false detection. The false detection rate is effectively zero for the least massive clouds because these false clouds, if they are exist, are too small to be recovered. We find that clouds with log$(M/M_{\odot}) \geq 5.9$ are well recovered by the program, while more than 80\% of clouds with log$(M/M_{\odot}) \approx 5.7$ are recovered (Figure 3). Therefore, we adopt the completeness level as log$(M/M_{\odot}) = 5.7$.

\subsubsection{Equal bin-width}

In the equal bin-width method, we group the masses into bins of equal width in log-space. Then, each histogram is fit with a straight line, weighted by the uncertainty of $dN/dM$, from the highest mass bin down to the completeness level. The slope of the best fit line, $x$, is the exponent of the mass function $dN/dM \propto M^{x}$.

We calculate the uncertainty on the number of clouds in each bin as follows. First, the uncertainties of the masses $\delta M$ are calculated through a bootstrap resampling method (RL06). Then, we use $\delta M$ to calculate the uncertainty in d$N$/d$M$ using Monte Carlo simulations. In these simulations, we resample the masses of the cloud, given a log-normal probability function with a mean $M$ and a standard deviation $\delta M$. The resampled masses are grouped into the same mass bins as the data, so that each simulation gives a new mass distribution. We repeat these steps 10,000 times, and take the uncertainty in d$N$/d$M$ as the standard deviation of these 10,000 simulations.

To check the robustness of our results, we vary the number of bins from 6 to 20. We find that $x$ varies from $-2.67$ to $-2.16$ (left panel of Figure 4). The uncertainty of the slope is taken from the covariance matrix of the fit. From these variations of $x$, we conclude that the equal bin-width method has large scatter due to the choice of the number of bins.

\subsubsection{Equal bin-size}

\citet{dagostino86} and \citet{maiz05} found that variable bin-widths with equally divided numbers of data points per bin can minimize the binning uncertainty, and hence, more robust. This is because no bin has a much smaller number of data points than the others, in contrast to the equal bin-width method. For our data, the actual number of data points in a bin is not exactly the same; it can differ from that in other bins by one data point, due to non integer numbers after division. To check the robustness of the result, we vary the number of bins as in the equal bin-width method, and fit the resulting histogram with a straight line. The results are indeed more robust than the equal bin-width method, with a maximum slope of $-2.31$ and a minimum slope of $-2.41$ (left panel of Figure 4).

\subsubsection{Cumulative distribution}

\begin{table}
\caption{Best-fit parameters of the cumulative mass distributions}
\centering
\begin{tabular}{ c c c c c }
\tableline
Region & Distance [pc] & $x$ & $M_0 [10^6 M_{\odot}]$ & $N_0$ \\[0.5ex]
\tableline
All & $\phantom{2}0 < d \leq 900$ & $-2.39 \pm 0.03$ & $4.12 \pm 0.08$ & $9.40 \pm \phantom{1}0.70$ \\[0.5ex]
Inner & $\phantom{2}0 < d \leq 170$ & $-1.40 \pm 1.19$ & $1.88 \pm 0.23$ & $9.13 \pm 33.21$ \\[0.5ex]
Ring & $170 < d \leq 580$ & $-2.38 \pm 0.03$ & $4.66 \pm 0.11$ & $5.66 \pm \phantom{1}0.47$ \\[0.5ex]
Outer & $580 < d \leq 900$ & $-2.46 \pm 0.12$ & $2.56 \pm 0.12$ & $4.67 \pm \phantom{1}1.12$ \\[0.5ex]
\tableline
\end{tabular}
\end{table}

In addition, we also calculate the mass function using the (truncated) cumulative distribution function \citep[e.g.][]{williams97,eros05b},
\begin{equation}
N(M'>M) = N_0 \left[ \left(\frac{M}{M_0}\right)^{x + 1} - 1 \right],
\end{equation}
and the simple (non-truncated) power-law distribution function,
\begin{equation}
N(M'>M) = \left(\frac{M}{M_0}\right)^{x + 1},
\end{equation}
where $M_0$ is the cut-off mass of the distribution and $N_0$ is the number of clouds with $M > 2^{1/(x+1)} M_0$, i.e. the cut-off point of the distribution. The truncated mass distribution takes into account that the mass distribution of a population lack clouds more massive than $M_0$. The cumulative distribution function is robust against the number of bins since we do not bin the data into a histogram.

We fit the cumulative distributions of masses using the orthogonal distance regression method in Scipy \citep{boggs90}. The fit is made for all data above the completeness level. We find $x = -2.39 \pm 0.03$ and $M_0 = (4.12 \pm 0.08) \times 10^6 M_{\odot}$, in agreement with the equal bin-size method. The data are inconsistent with simple (non-truncated) power-law mass distributions (middle panel of Figure 4). We find that the GMC mass distribution in NGC4526 is steeper than in the inner MW \citep[$x = -1.5$;][]{eros05b}, but comparable to the GMC mass distribution in the outer MW \citep[$x = -2.1$;][]{eros05b} and central M33 \citep[$x = -2.0$;][]{erosetal07}.

All three methods of measurements suggest that $x < -2$. In this case, most of the mass resides in the low mass clouds. Furthermore, the total mass diverges for integration down to an infinitely small mass. Hence, there must be a lower limit to the cloud masses or a change in the slope, i.e. $x > -2$ for lower mass clouds below our completeness level, so that the total mass remains finite.

We further divide the galaxy into three distinct regions: inner ($0 < d \leq 170$ pc), molecular ring ($170 < d \leq 580$ pc), and outer ($580 < d \leq 900$ pc) region (concentric ellipses in Figure 1), and measure their mass distributions. The molecular ring and the outer region have a similar mass distributions with an exponent of $x \approx -2.4$, while the inner region is much flatter ($x \approx -1.4$), albeit with a large uncertainty due to the small number of clouds in the inner region. The best-fit parameters are compiled in Table 2. A radial dependence of the mass function was also discovered in M33 \citep{gratier12} and M51 \citep{colombo14}.

We calculate the total mass of detected GMCs (including the non-resolved clouds) to be $M_{\rm GMC} = (2.0 \pm 0.1) \times 10^8 M_{\odot}$. The total H$_2$ mass is $M({\rm H}_2) = (3.8 \pm 1.1) \times 10^8 M_{\odot}$ \citep{young08}\footnote{We recalculate the total H$_2$ mass using $X_{\rm CO}$ as in this paper.}, so the fraction of molecular mass that residing in GMCs is $M_{\rm GMC}/M({\rm H}_2) \approx 0.53$. This value is formally a lower limit, since there are GMCs with masses below the completeness level of our observations that are undetected. The rest of the molecular gas may be in the form of diffuse gas which is undetected by interferometric observations.

\subsection{Larson's Relations}

\citet{larson81} found that the velocity dispersion of GMCs is correlated with their size through a power-law relation with exponent of $\sim 0.3$. This correlation is similar to that expected if turbulence governs the velocity dispersion within clouds as described by the Kolmogorov law. In subsequent work, S87 refined the exponent to be $0.5 \pm 0.05$ for GMCs in the Milky Way's inner disk.

\begin{figure*}
%\figurenum{text}
\epsscale{1.2}
\plotone{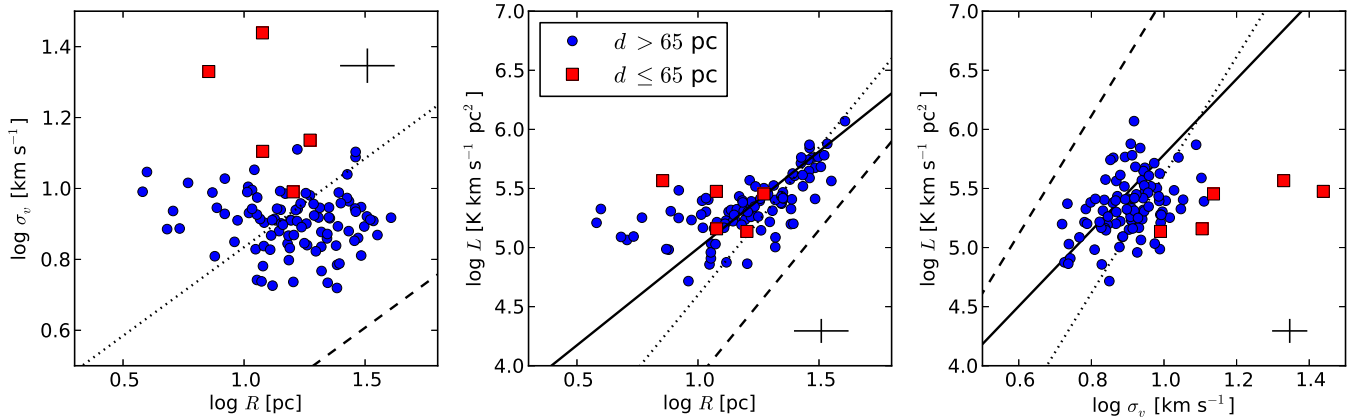}
\label{fig:5}
%\plottwo{epsfile}{epsfile}
\caption{GMCs properties (radius, velocity dispersion, and luminosity) are plotted relative to one another. The color codes are for inner GMCs (distance $\leq 65$ pc; red squares) and outer GMCs (distance $> 65$ pc; blue dots). The fits of all data points (including the inner clouds) are shown as solid lines. The typical uncertainty is shown as a cross sign in the corner of each panel. For clarity, we do not plot the uncertainties of individual data. However, we fit the data points by including the non-uniform error bar of the individual data point, not only the typical uncertainty. The dashed lines are Larson's relations for the Milky Way disk (S87), and the dotted lines are Larson's relations with different normalization factors: 3, 0.03, and 5 from left to right panel, respectively. It shows that GMCs in NGC4526 are more turbulent and more luminous than equal-size clouds in the Milky Way disk. There is no size-linewidth relation, in contradiction to the expectation from Larson's relation (left panel).}
\end{figure*}

Larson's relations in the Milky Way consist of two independent equations \citep[e.g. S87,][]{bolatto08}:
\begin{equation}
\sigma_v \approx 0.72 \left(\frac{R}{\rm pc}\right)^{0.5} \ {\rm km \ s}^{-1}
\end{equation}
and either
\begin{equation}
L_{\rm CO} \approx 25 \left(\frac{R}{\rm pc}\right)^{2.5} \ {\rm K \ km \ s}^{-1} \ {\rm pc}^2
\end{equation}
or
\begin{equation}
L_{\rm CO} \approx 130 \left(\frac{\sigma_v}{{\rm km \ s}^{-1}}\right)^5 \ {\rm K \ km \ s}^{-1} \ {\rm pc}^2.
\end{equation}
Since the linewidth, $\Delta V$, is just $\Delta V = \sigma_v \sqrt{8 \ {\rm ln} \ 2}$, we refer to equation (8) as the size-linewidth relation. For extragalactic clouds in the Local Group, Bolatto et al. (2008) found $\sigma_v \propto R^{0.6}$, $L_{\rm CO} \propto R^{2.54}$, and $L_{\rm CO} \propto \sigma_v^{3.35}$, which is close to the Milky Way relations.

Interestingly, we find no size-linewidth correlation for NGC4526 (left panel of Figure 5), which is parameterized by very weak Pearson and Spearman correlation coefficients ($r_{\rm ps} = -0.18$ and $r_{\rm sp} = -0.14$). This result is in line with GMCs in M33 \citep{gratier12} and M51 \citep{colombo14}, where no clear trend was observed ($r_{\rm sp} = 0.12$ and 0.16 for M33 and M51, respectively). The NGC4526 data are located above the Milky Way's relation, which means that for a given size, GMCs in the bulge of NGC4526 have a higher velocity dispersion than those in Milky Way disk GMCs by a factor of $\sim 3$. This could be due to an environmental effect, since \citet{shetty12} and \citet{colombo14} found evidence that GMCs in the central regions of the Milky Way and M51 have a higher velocity dispersion than those in the disks. We discuss this environmental effect in $\S 5.4$.

We also plot cloud luminosity against velocity dispersion and size in the middle and right panels of Figure 5. The results of the error-weighted fit are
\begin{equation}
L_{\rm CO} = 2258.5_{- 708.8}^{+ 1033.0} \left(\frac{R}{\rm pc}\right)^{1.6 \pm 0.1} \ {\rm K \ km \ s}^{-1} \ {\rm pc}^2
\end{equation}
and
\begin{equation}
L_{\rm CO} = 381.3_{- 238.4}^{+ 635.8} \left(\frac{\sigma_v}{{\rm km \ s}^{-1}}\right)^{3.2 \pm 0.5} \ {\rm K \ km \ s}^{-1} \ {\rm pc}^2,
\end{equation}
which is shallower than the exponents in the Milky Way relations (equations 9 and 10). The correlation coefficients are moderate for the luminosity-size relation ($r_{\rm ps} = 0.63$ and $r_{\rm sp} = 0.67$), and weak for the luminosity-velocity dispersion relation ($r_{\rm ps} = 0.33$ and $r_{\rm sp} = 0.36$).

The quoted results above take into account all resolved clouds. In Figure 5, the inner clouds with distance $\leq 65$ pc from the galactic center tends to have higher velocity dispersion, which may be due to contamination by the galactic shear (see $\S 4$ for detailed discussion of cloud kinematics). The conclusions for the size-linewidth and size-luminosity relation are not affected if we exclude these inner clouds. However, the slope of linewidth-luminosity relation become steeper by excluding those inner clouds ($L_{\rm CO} \propto \sigma_v^{3.8 \pm 0.6}$).

In the middle panel of Figure 5, the clouds in NGC4526 lie below the Milky Way luminosity-velocity dispersion relation. Thus, for a given velocity dispersion, the clouds are less luminous than GMCs in the disk of the Milky Way. Because CO luminosity is a tracer of the amount of molecular gas, the clouds in NGC4526 are more turbulent per unit mass than those in the Milky Way. Also, from the right panel of Figure 5, clouds in NGC4526 are $\sim 5$ times more luminous than equal size clouds in the Milky Way, which means they have a higher surface density. This result is in agreement with GMCs in M51 \citep{colombo14}, where GMCs in the central region are brighter than those in the inter-arms region. Again, this could be due to environmental effects which are discussed in $\S 5.4$.

Finally, we find no bias that affects the results due to the choice of input parameters of the CPROPS program. Even though the properties of individual clouds vary by changing the input parameters, the overall distributions are similar (see Appendix D for details).

\section{Cloud Kinematics}

\subsection{Velocity Gradient of the Cloud}

\begin{figure}
%\figurenum{text}
\epsscale{1.2}
\plotone{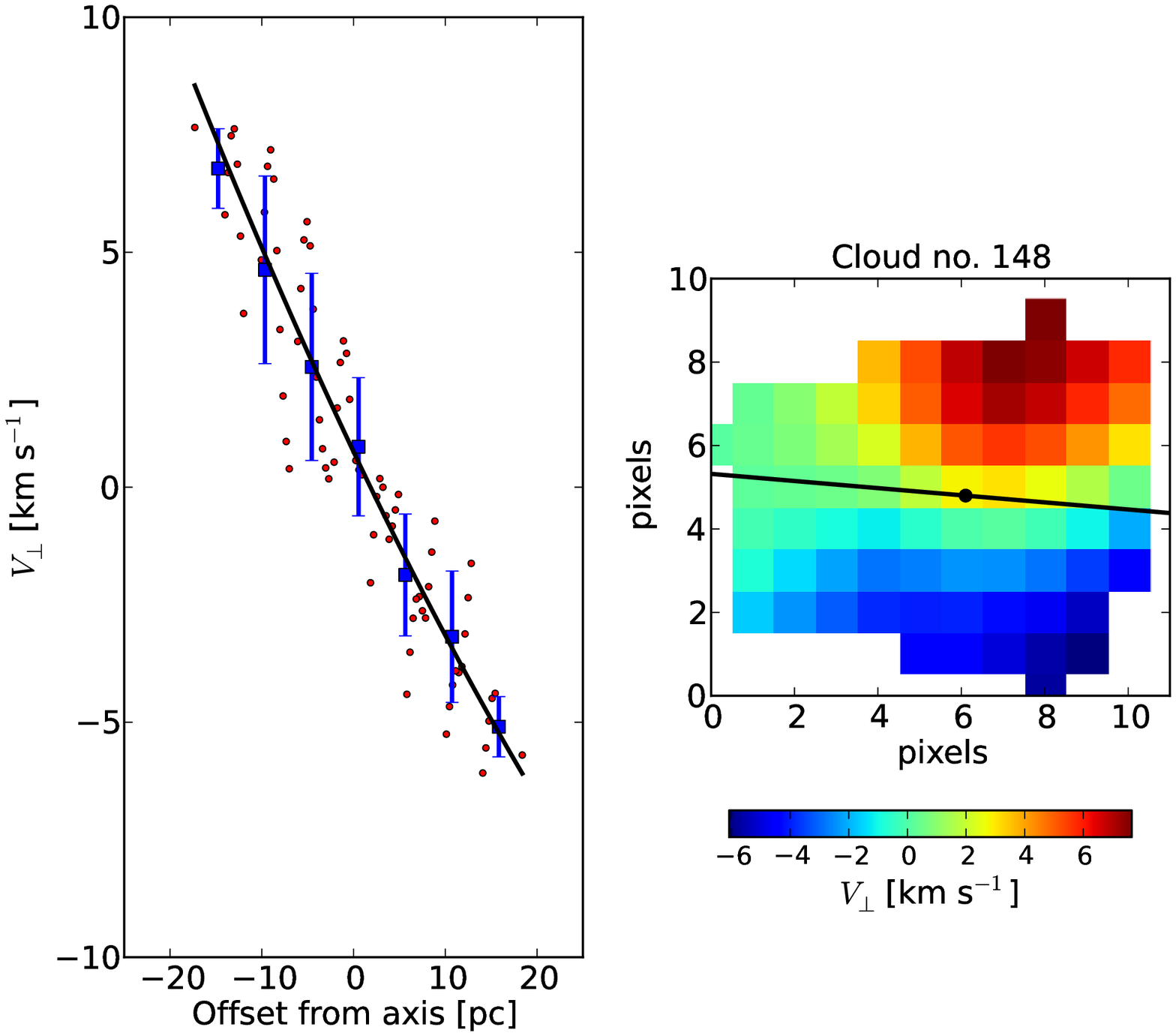}
\epsscale{1.2}
\plotone{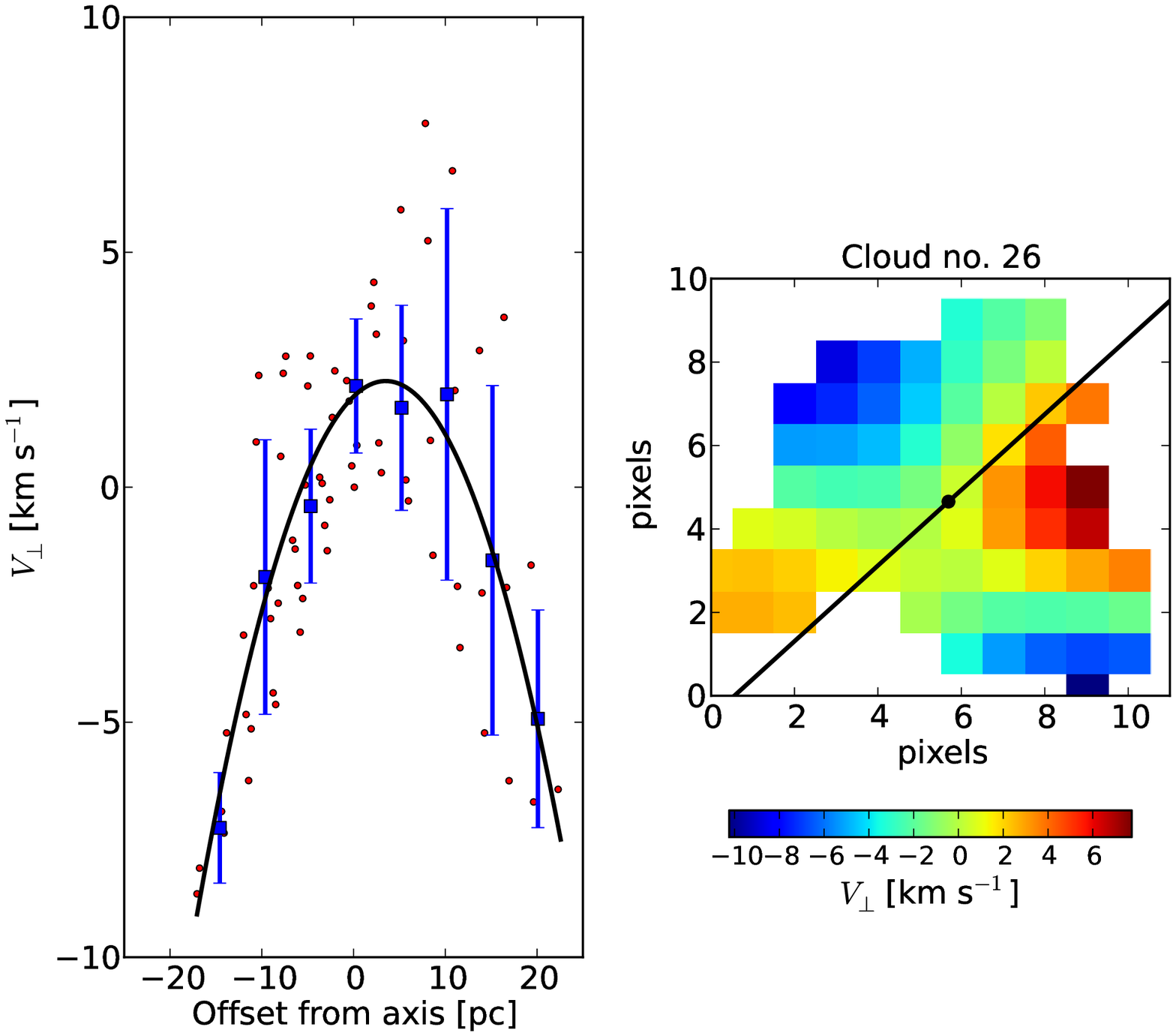}
\label{fig:6}
%\plottwo{epsfile}{epsfile}
\caption{Examples of plane fitting to find the rotation signature of GMCs. The right panels are GMC first-moment maps with the rotation axis (black line) overplotted. On the left panels, the mean velocity of each pixel is plotted against its perpendicular distance from the rotation axis. The black line is the fit and blue dots are the mean of the velocity in bins of perpendicular distance from the rotation axis (i.e. $v_{\bot}$ vs $d$). The cloud in the top panels shows the signature of solid body rotation with $\chi_{\nu, \rm{line}}^2 \approx 0.18$, while the cloud in the bottom panels shows bow-shock motions and is well fitted by a parabolic curve with $\chi_{\nu,{\rm para}}^2 \approx 0.68$.}
\end{figure}

Previous studies \citep[e.g.][]{imara11,eros03,phillips99,kane97,goodman93} have found velocity gradients across atomic and molecular clouds and clumps, which are interpreted as rotation of the GMCs. Moreover, most of the GMC rotation exhibits solid body rotation. Here, we perform analyses on the resolved clouds to gain insight into the origin of their kinematics and the role of rotation in the dynamical stability of GMCs.

\begin{figure*}
%\figurenum{text}
\epsscale{1.2}
\plotone{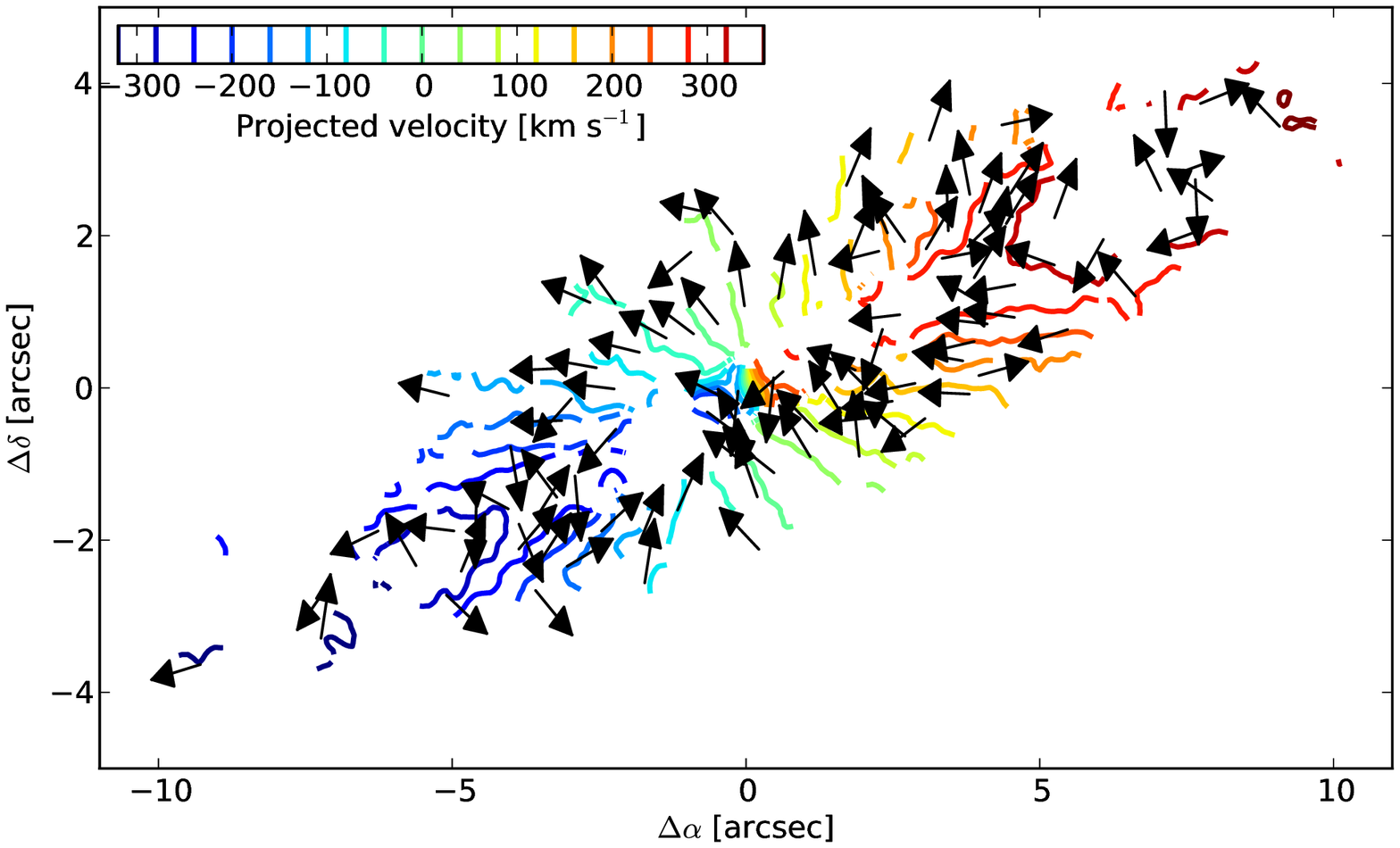}
\label{fig:7}
%\plottwo{epsfile}{epsfile}
\caption{The angular momentum vectors of GMCs (black arrows), overplotted with the isovelocity contours of NGC4526 (color coded by their projected velocities) convolved with a Gaussian kernel. There is a strong tendency for the vectors to be tangential to the isovelocity contours, as expected if the measured velocity gradients of GMCs are just the projection of the galaxy rotation. Correlations between the angular momentum of the clouds and isovelocity contours of the galaxy are shown in Figure 8.}
\end{figure*}

In order to quantify any rotation signature, we do the following. First, the velocity field (first-moment map) of the cloud is smoothed with a Gaussian kernel, where the dispersion of the Gaussian kernel is half the telescope beamwidth. The aim of this smoothing is to `average` the velocity field at the cost of losing independence among the neighboring pixels (left panels of Figure 6). Then, we fit the first-moment map of individual clouds with a plane \citep[e.g.][]{goodman93,eros03,imara11}:
\begin{equation}
v_{\rm los} = v_0 + a (x-x_0) + b (y-y_0),
\end{equation}
where $(x_0,y_0)$ is the cloud's central pixel coordinate, and
\begin{equation}
a = \frac{\partial v}{\partial x} \ \ {\rm and} \ \ b = \frac{\partial v}{\partial y}
\end{equation}
are the velocity gradients along the $x$ and $y$ axes. $v_0$, $a$, and $b$ are free parameters to be determined from the fit.

The angle from the positive $x$-axis to the receding part of kinematical major axis of the cloud is ${\rm tan}^{-1} (b/a)$, and hence, the angle to the cloud rotation axis (i.e. the angular momentum vector) is $\theta = {\rm tan}^{-1} (b/a) + 90^{\circ}$ (i.e., right hand rule).

The angular speed of the cloud is given by
\begin{equation}
\Omega \ {\rm cos}(\psi) = \sqrt{ \left( \frac{\partial v}{\partial x} \right)^2 + \left( \frac{\partial v}{\partial y} \right)^2},
\end{equation}
where $\psi$ is the angle from the cloud rotation axis to the sky plane. Since we can not measure $\psi$ directly, we drop the cos$(\psi)$ term from equation (15). Thus, the true angular speed is underestimated by a factor of cos$(\psi)$, i.e. $\Omega_{\rm projected} = \Omega_{\rm true} \ {\rm cos}(\psi)$.

The next step is to check whether the clouds show solid-body or differential rotation. We plot the mean velocity of each pixel within a cloud against its perpendicular distance from the cloud rotation axis, i.e. $v_{\bot}$ vs. $d$ (Figure 6), and then we fit the data with a straight line. Solid body rotators should show a clear linear behavior on this plot, where the constant slope is the angular speed, i.e. $v_{\bot} = \Omega \ d$. On the other hand, the slope of Keplerian rotators varies with distance from the center, i.e. the slope gets shallower outside (resembles an S-shape; $v_{\bot} \propto d^{-0.5}$), while bow-shock motions tend to have a parabolic shape \citep{kane97}.

Finally, we divide the clouds into two groups: clouds that show solid-body rotation (SB) and clouds that show deviations from solid-body rotation (NSB). This division is based on the reduced $\chi^2$ value from the binned $v_{\bot}$ (blue squares in Figure 6) to the straight line fit. We classify clouds with $\chi_{\nu, {\rm line}}^2 \leq 1.5$ as SB and the rest as NSB. For NSB, we also fit $v_{\bot}(d)$ with a parabolic curve and calculate its $\chi_{\nu, {\rm para}}^2$. Eye inspections confirm that this classification is reasonable. As a result, 46 of 103 resolved clouds are classified as SB, while the rest are NSB. Examples of SB and NSB clouds are shown in Figure 6.

\subsection{Origin of Velocity Gradients}

\begin{figure*}
%\figurenum{text}
\epsscale{1}
\plotone{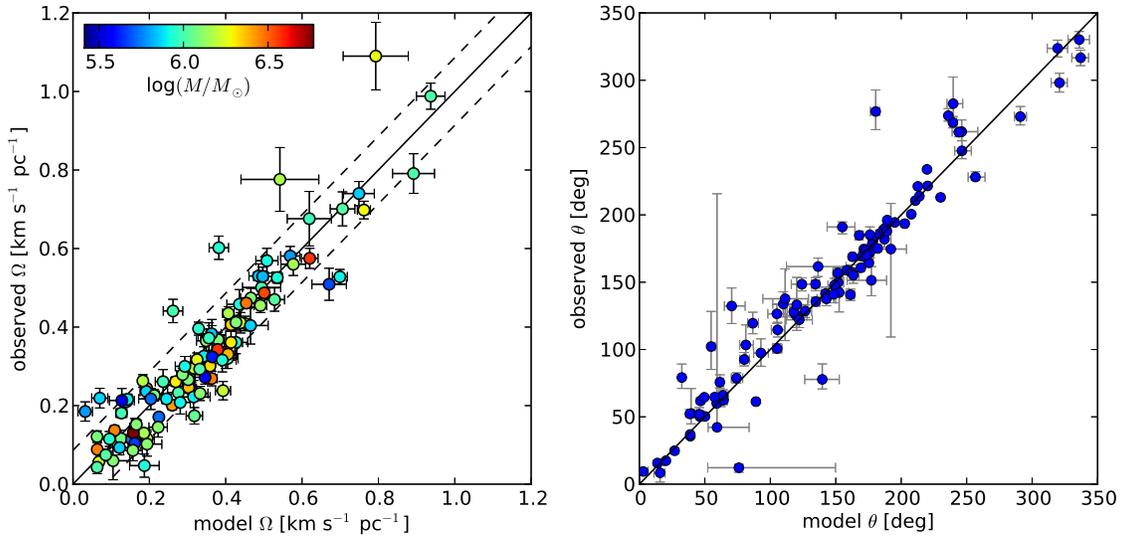}
\label{fig:8}
%\plottwo{epsfile}{epsfile}
\caption{Correlations between the model and observed angular speed $\Omega$ (left panel) and the angle of rotation axis $\theta$ (right panel). The observed $\Omega$ and $\theta$ are calculated from the best fit of the velocity field of the cloud \citep[e.g.,][]{goodman93}. The model $\Omega$ and $\theta$ are purely based on the galaxy rotation and created using the KinMS package of \citet{davis13} as described in the text. The error bars are derived from the covariance matrix of the best fit. The excellent one-to-one correlations (solid line) indicate that the velocity gradients that we measure are actually just a projection of the galaxy rotation. The dashed lines on the left panel are the standard deviation from Monte Carlo simulations that gives the upper limit of the angular speed of the cloud.}
\end{figure*}

A purely rotating galaxy with an inclination angle $i$ has a line of sight velocity component of $V_{\rm los} = V_{\rm sys} + V(R) \ {\rm cos}(\phi) \ {\rm sin}(i)$ at ($R,\phi$), where $V_{\rm sys}$ is the systemic velocity of the galaxy, $V(R)$ is the circular velocity at radius $R$ from the galactic center, and $\phi$ is the angle from the kinematic major axis of the galaxy. In the simplest case, for a region with a flat rotation curve, i.e. $V(R) = {\rm constant}$, the observed isovelocity contours of the galaxy are given by contours of equal $\phi$. Therefore, velocity gradients exist along any path that perpendicularly crossing those isovelocity contours. The last statement is true for any rotation curve, not just for a flat rotation curve.

On the other hand, we also find velocity gradient in any small patches of the data that occupy the GMC regions. This velocity gradient can be due to a projection of the galaxy rotation, and therefore, can mimic the cloud rotation. In Figure 7, we show the angular momentum vectors of GMCs, overplotted with the isovelocity contours of NGC4526. The tendency of the angular momentum vectors of the clouds to be tangential to the isovelocity contours of the galaxy suggests that the velocity gradients of the clouds are actually just a projection of the galaxy rotation.

In order to quantify our finding, we create a gas dynamical model using the KinMS (Kinematic Molecular Simulation) package of \citet{davis13}. This model is basically a purely rotating disk based on a rotation curve of the galaxy, i.e. this is what the galaxy look like if its dynamic is just a rotation. The rotation curve is calculated from the multi-Gaussian expansion (MGE) of \citet{cappellari02} to the {\it I}-band images of HST and MDM 1.3-m telescope, and includes the presence of a supermassive black hole at the center \citep{davis13b}. This MGE fit gives the mass distribution of the galaxy, and hence the galaxy rotation curve, parameterized by the stellar mass-to-light ratio and the galaxy inclination. The model is inclined so that it matches the inclination of the observed galaxy ($i \approx 79^{\circ}$). Any deviations of the data from the model can be caused by small scale turbulence, inflow or outflow gas motion, and the cloud rotation.

Then, we measure the angular momentum of the model at the location of the observed clouds by using the same method as described in $\S 4.1$. We find a strong one-to-one correlation for both angular speed and rotation angle between the model and the data, albeit with small scatter (Figure 8). The correlation holds true for both SB and NSB groups. This reinforces our inference that the velocity gradients of the clouds are just a consequence of the underlying velocity field due to galaxy rotation, i.e. the gas within clouds moves following the galaxy rotation. Therefore, $\Omega$ that is calculated using the plane-fitting method (equation 15) is not the intrinsic angular speed of the cloud.

In this case, the gas within the cloud must rotate due to galactic shear. The amount of shear is given by the Oort constant $A$ \citep{fleck81}:
\begin{equation}
\Omega_{\rm shear} = \left| \frac{\Delta v}{\Delta r} \right| = |2A| = \left| \frac{V_0}{R_0} - \left( \frac{dV}{dR} \right)_0 \right|,
\end{equation}
where the subscripts 0 denote the evaluation at the location of the GMCs, and $V(R)$ is the rotation curve of the galaxy. Hereafter, {\it we take this shear as the angular speed of the cloud}, not $\Omega$ from equation (15). Furthermore, the angle of the rotation axis $\theta$ from the plane-fitting is not physically meaningful anymore, because if the cloud rotation is due to galactic shear then their rotation axes tend to be parallel to the galaxy rotation axis.

The intrinsic scatters in Figure 8 may be due to the intrinsic angular momentum of the cloud, $\Omega_{\rm cloud}$, that is not originated from the galactic shear. From 1,000 Monte Carlo simulations, we determine the upper limit of the cloud's angular speed to be $\Omega_{\rm cloud} < 8.6 \times 10^{-2}$ km s$^{-1}$ pc$^{-1}$ (dashed line in Figure 8), which is comparable to the angular speed of GMCs and HI cloud in M33 \citep{eros03,imara11}. This upper limit is generally smaller than the galactic shear at the cloud's location, i.e. $\Omega_{\rm cloud} < \Omega_{\rm shear}$, so that $\Omega \approx \Omega_{\rm shear}$. No correlation between $\Omega_{\rm shear}$ and the cloud's mass is found (Figure 8).

Based on our analysis, any measurement of the velocity gradient of an extragalactic cloud must be performed carefully, to avoid bias due to the projection of the galaxy rotation. The only exception is if the galaxy is nearly face-on ($i \approx 0^{\circ}$), as the line of sight velocity due to galaxy rotation is negligible.

\subsection{Stability of Rotating Clouds}

\begin{figure}
%\figurenum{text}
\epsscale{1.2}
\plotone{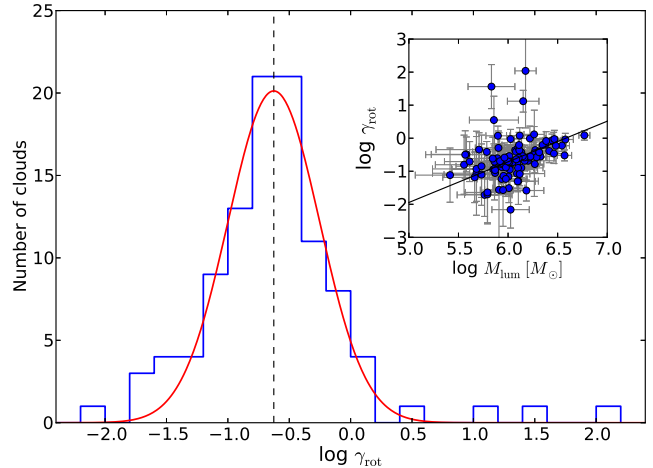}
\label{fig:9}
%\plottwo{epsfile}{epsfile}
\caption{Distribution of $\gamma_{\rm rot}$, defined as the ratio between rotational and turbulent energy. The dashed line is the mean of the log-normal fit (red). For the majority of the clouds, the rotational energy due to galactic shear is smaller than the turbulent energy. From the correlation between $\gamma_{\rm rot}$ and luminous mass (shown as an inset), we infer that the relative importance of rotation over turbulence is increasing for more massive clouds.}
\end{figure}

\begin{figure}
%\figurenum{text}
\epsscale{1.2}
\plotone{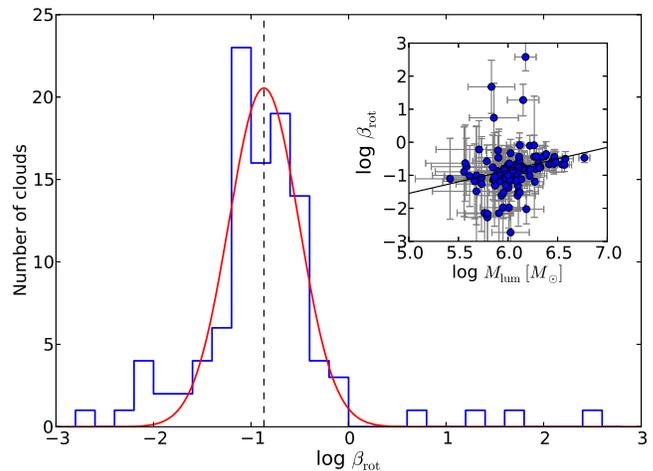}
\label{fig:10}
%\plottwo{epsfile}{epsfile}
\caption{Same as Figure 9, but for $\beta_{\rm rot}$, which is defined as the ratio between rotational and gravitational energy. Most of the clouds are not in rotational equilibrium, i.e. $\beta_{\rm rot} < 1$. The four clouds in the tail of the distribution are the innermost clouds that suffer strong shear. From the correlation between $\beta_{\rm rot}$ and luminous mass (shown as an inset), we infer that the relative importance of rotation over gravity is increasing for more massive clouds.}
\end{figure}

In $\S 3.2$, we assumed that the clouds were supported by turbulence only. In the presence of rotation (due to shear), the rotational energy also contributes to the clouds' stability against gravitational collapse. Here, we define the parameter $\gamma_{\rm rot}$ as the ratio of rotational over turbulent energy:
\begin{equation}
\gamma_{\rm rot} = \frac{p \Omega^2 R^2}{3 \sigma_v^2},
\end{equation}
where $p = 2/5$, the value for a uniform sphere \citep{goodman93}.

In Figure 9, we show the histogram of the $\gamma_{\rm rot}$ values of the clouds. A log-normal fit to the distribution yields a mean $\gamma_{\rm rot} \approx 0.24$ with a standard deviation of 0.37 dex. Roughly 92\% of the resolved clouds have $\gamma_{\rm rot} < 1$. This means that the rotational energy is smaller than the turbulent energy for the vast majority of the clouds. There is also a correlation with mass as $\gamma_{\rm rot} \propto M_{\rm lum}^{1.23 \pm 0.16}$, so that the relative importance of rotation over turbulence is increasing for more massive clouds.

It is also useful to define the ratio between rotational kinetic energy and self-gravitational energy:
\begin{equation}
\beta_{\rm rot} = \frac{1}{2} \frac{p}{q} \frac{\Omega^2 R^3}{GM},
\end{equation}
where $p/q = 2/3$, the value for a uniform sphere \citep{goodman93}. Clouds with $\beta_{\rm rot} \approx 1$ are rotationally-stable against gravitational collapse. About 96\% percent of the resolved clouds have $\beta < 1$. A log-normal fit to the distribution yields a mean $\beta_{\rm rot} \approx 0.14$ with a standard deviation of 0.36 dex (Figure 10). This means that the rotational energy is smaller than the gravitational energy for the vast majority of the clouds. Furthermore, there is a correlation with mass as $\beta_{\rm rot} \propto M_{\rm lum}^{0.70 \pm 0.21}$, so that the relative importance of rotation over gravity is increasing for more massive clouds.

Finally, the virial parameter $\alpha$, which includes turbulence, gravity, and rotation, can be expressed as
\begin{equation}
\alpha = \beta_{\rm rot} \left( 1 + \frac{2}{\gamma_{\rm rot}} \right).
\end{equation}
Non-magnetic, rotating clouds with $\alpha \approx 1.13$ are in virial-equilibrium (BM92), while clouds with $\alpha \gg 1$ are either pressure-confined clouds, or still gravitationally bound but with an underestimated CO-to-H$_2$ conversion factor.

\begin{figure}
%\figurenum{text}
\epsscale{1.2}
\plotone{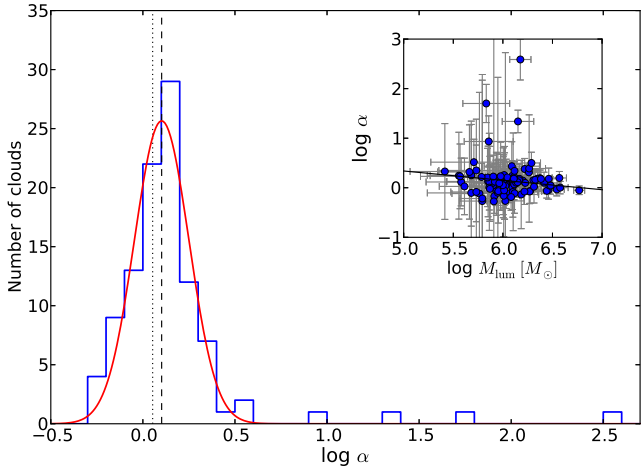}
\label{fig:11}
%\plottwo{epsfile}{epsfile}
\caption{Same as Figure 9, but for the virial parameter $\alpha$. The dashed line is the mean Gaussian fit, and the dotted line is $\alpha = 1.13$. The four clouds in the tail of the distribution are the innermost clouds that undergo a strong shear. Except for those four clouds, the cloud population is in gravitational equilibrium, i.e. the mean $\alpha \approx 1.26$. The inset shows anti-correlation between the virial parameter and the luminous mass.}
\end{figure}

The distribution of $\alpha$ values is shown in Figure 11, where it can be approximated by a log-normal distribution with a tail at the high end. The log-normal mean is $\alpha \approx 1.26$ with a standard deviation of 0.15 dex. This suggests that the GMC population in NGC4526 is gravitationally-bound, even after the inclusion of rotational energy. If we define non-gravitational equilibrium as having $\alpha > 3.5 \approx 0.5$ dex (i.e. the tail of the distribution), then only $\approx$ 4\% of resolved clouds are not gravitationally-bound. As shown in the inset in Figure 11, by excluding those four clouds, $\alpha$ has a shallow correlation with mass as $\alpha \propto M_{\rm lum}^{-0.19 \pm 0.05}$.

Further investigation reveals that the four gravitationally-unbound clouds (with log $\alpha > 0.5$) are the clouds closest to the center of the galaxy, at a distance of $\approx 10$, 34, 42, and 54 pc. These clouds suffer strong shear ($|\Delta v/ \Delta r| \sim 10$ km s$^{-1}$ pc$^{-1}$) due to the presence of a supermassive black hole (SMBH) at the center of NGC4526, with $M_{\rm BH} \approx 4.5 \times 10^8 M_{\odot}$ and radius of influence of $\approx 45$ pc \citep{davis13b}. This SMBH makes the circular velocity curve of the galaxy increases abruptly towards the galactic center (Figure 12), and hence yields a large Oort $A$ constant.

\section{Discussion}

\subsection{Pressure Balance}

\begin{figure}
%\figurenum{text}
\epsscale{1.2}
\plotone{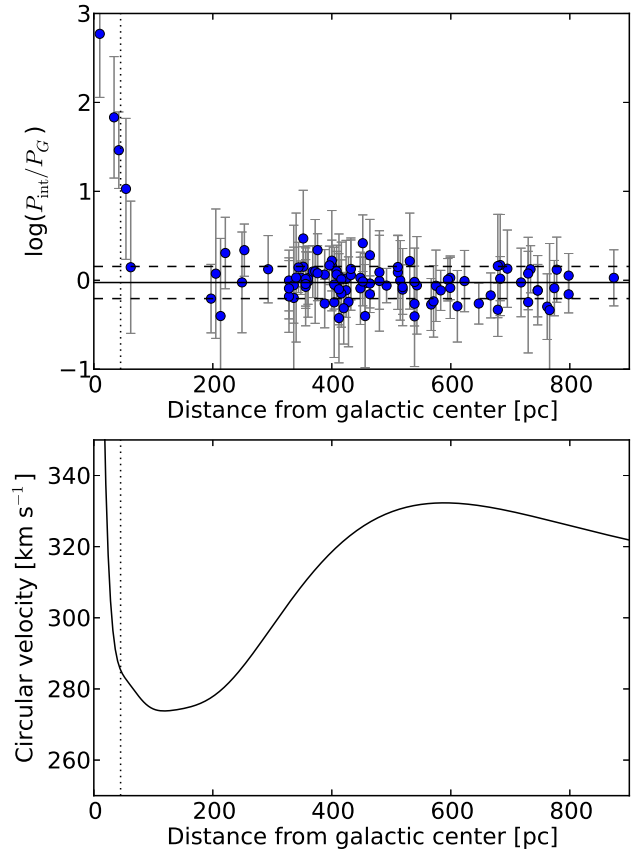}
\label{fig:12}
%\plottwo{epsfile}{epsfile}
\caption{Top: ratio of internal and gravitational pressure as a function of distance from the galactic center. Except for the four innermost clouds, the data points are consistent with gravitationally-bound clouds. The $1\sigma$ scatters are indicated with dashed lines. The SMBH radius of influence ($\approx 45$ pc) are indicated as vertical dotted lines. Bottom: galaxy circular velocity curve. The sharp increase near the center is due to the presence of a supermassive black hole \citep{davis13b}.}
\end{figure}

In general, the dynamical equilibrium state of a cloud can be written as
\begin{equation}
P_{\rm int} + P_B = P_G + P_{\rm ext},
\end{equation}
where $P_{\rm int} \approx P_{\rm turb} \ (1 + \gamma_{\rm rot})$ is the internal pressure of the cloud, including the correction factor $(1 + \gamma_{\rm rot})$, due to the contribution of rotation. $P_{\rm turb} = \bar{\rho}\sigma_v^2$ is the kinetic pressure due to turbulence, $P_{\rm ext}$ is the external pressure of the ambient medium, and $P_B = B^2/8\pi$ is the magnetic pressure. Here, we assume that the thermal pressure is much smaller than the pressure due to turbulent motion, and it is neglected. $P_G$ is the internal gas pressure that is required to support the cloud against gravity in the absence of any other forces (BM92):
\begin{equation}
\frac{P_G}{k} = 1.3 \ \bar{\phi_G} \left(\frac{M}{M_{\odot}}\right)^2 \left(\frac{R}{\rm pc}\right)^{-4} \ {\rm K \ cm}^{-3},
\end{equation}
where $\bar{\phi_G}$ is a dimensionless factor that measures the ratio between the gravitational pressure of ellipsoidal and spherical clouds, and depends only on the cloud's axis ratio: $\sigma_{\rm maj}/\sigma_{\rm min}$ for prolate clouds and $\sigma_{\rm min}/\sigma_{\rm maj}$ for oblate clouds. Here, we assume all clouds are prolate. The value of $<\phi_G>$ for oblate clouds is within the uncertainty of that for prolate clouds.

For non-magnetic, rotating, gravitationally-bound clouds: $P_{\rm int} \approx P_G$. In Figure 12, we plot log$(P_{\rm int}/P_G)$ versus the distance of clouds from the galactic center. Except for the four innermost clouds, log$(P_{\rm int}/P_G)$ has a mean value of $\approx -0.03$ and standard deviation of $\approx 0.18$ dex, consistent with a gravitationally bound state. We do not include the error bars of individual data point to calculate those values.

As we mentioned in $\S 4.3$, there are four central clouds that experience strong galactic shear. By using equation (20) and an assumption of zero magnetic pressure, the external pressure that is required to bind the clouds against galactic shear is $P_{\rm ext} \sim 10^9$ K cm$^{-3}$, which is extremely high and unrealistic. We argue that these are unbound clouds. If nothing balances the shear, then these clouds will be ripped apart by strong shear within a timescale of $\sim 2\pi/|\Delta v/\Delta r| \sim$ 1 Myr. This timescale is smaller than the expected lifetimes of GMCs \citep[$\approx$ 30 Myr;][]{blitz80} based on the clumpiness nature of GMCs and destruction processes from massive star formation inside GMC complexes. Another evidence of the role of shear in the destruction of GMCs is found in the M51 disk, where \citet{miyamoto14} reported that the locations of giant molecular associations are anti-correlated with the shear strength.

\subsection{Does a Size-linewidth Relation Really Exist?}

\begin{figure}
%\figurenum{text}
\epsscale{1.2}
\plotone{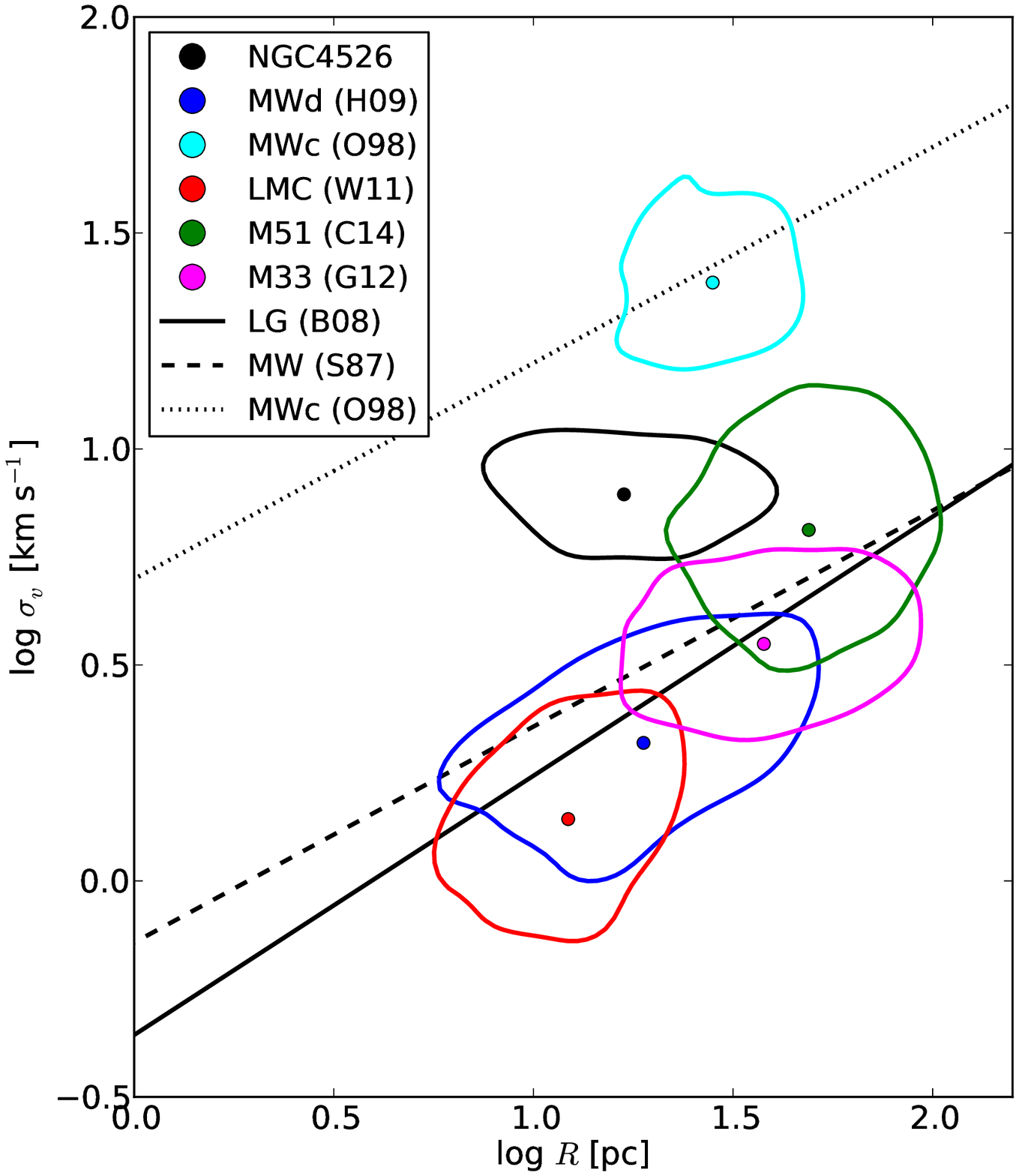}
\label{fig:13}
%\plottwo{epsfile}{epsfile}
\caption{Plots of size vs. velocity dispersion for extragalactic GMC populations. The contours enclose 68\% of the distribution of data points of a given galaxy: NGC4526 (black; this paper), Milky Way disk \citep[blue;][]{heyer09}, Milky Way center \citep[cyan;][]{oka98}, Large Magellanic Cloud \citep[red;][]{wong11}, M51 \citep[green;][]{colombo14}, and M33 \citep[magenta;][]{gratier12}. The centers of a Gaussian fit to each distribution are shown as filled circles. The dashed line is the Milky Way disk relationship (S87), the dotted line is the Milky Way center relationship \citep{oka98}, and the solid line is the Local Group relationship \citep{bolatto08}.}
\end{figure}

Clouds in NGC4526 do not show a size-linewidth relation (see Figure 5), in contrast to the previous arguments that supported the Larson's `law` \citep[e.g. in the Milky Way disk; S87, and Local Group galaxies;][]{bolatto08}. Most of the Local Group members are late-type galaxies, while our clouds reside in the bulge of a S0-type galaxy. Does this discrepancy suggest that GMC properties in the early-type galaxies are intrinsically different? Here, we argue that the size-linewidth relation may not exist in all galaxy morphologies, because the cloud's size and linewidth within a single galaxy only have weak to modest correlation coefficients.

Recent studies of GMCs in spirals, such as M33 \citep{gratier12} and M51 \citep{colombo14}, also found no clear size-linewidth relation, with a Spearman correlation coefficient $r_{\rm sp}$ of 0.12 and 0.16, respectively. A modest correlation ($r_{\rm sp} = 0.51$) was found by \citet{heyer09}, who re-examined S87 clouds in the Milky Way disk using more sensitive instruments. Furthermore, clouds in the LMC also show a weak correlation \citep[$r_{\rm sp} = 0.37$;][]{hughes13,wong11}. These evidence suggest that the size-linewidth relation may not exist in all galaxy morphologies.

If the argument for a size-linewidth relation is not conclusive for GMCs {\it within} a single galaxy, then how about a compilation of GMC data from various galaxies \citep[e.g.][]{bolatto08}? In this case, one must pay attention to different data sets that have different physical resolutions and sensitivities. Coarse resolution and low S/N observations can only measure average properties within a larger area, without the ability to decompose the CO structure into multiple smaller clouds, while finer resolution observations tend to over-decompose CO emission into smaller scale structures. This means that the identified GMCs in different data sets are likely to probe different scales of CO emission. This bias, which is due to the ability to decompose structure in GMCs, is separate from the bias of measured properties due to finite resolution and sensitivity, which has been minimized by the CPROPS program. Hence, plotting those data in a size-linewidth diagram leads us to compare different structure of GMCs. For example, \citet{bolatto08} compare composite extragalactic GMCs that have been observed with a range of resolution from $\sim 6$ pc (about the size of a clump) to $\sim 117$ pc (about the size of a giant molecular association). Therefore, any scatter in the size vs. linewidth diagram is overcome by the large range of GMC size, which gives rise to a slope. Even in this case, however, the correlation coefficient is still moderate ($r_{\rm sp} = 0.57$).

A self-consistent study of the size-linewidth relation, then, requires a common physical resolution and sensitivity across the extragalactic GMC data sets. Any similarity or discrepancy among the extragalactic GMCs measured in this way would then be genuine. In a recent work, \citet{hughes13} showed that a size-linewidth relation is apparent when M51, M33, and LMC data were analyzed at their original (different) resolutions and sensitivities \citep[as in][]{bolatto08}, but no compelling evidence was found when the data were degraded to a single (`matched`) common resolution and sensitivity. From the `matched` data, they infered that GMCs in M51 are in general larger, brighter, and have higher velocity dispersions than equivalent structures in M33 and the LMC, which can be interpreted as a genuine variation of GMC properties.

In this respect, we can compare NGC4526 and LMC data \citep{wong11}, since they have comparable physical resolutions and sensitivities ($\sim 20$ pc and $\sigma_{\rm rms} \approx 0.7$ K for NGC4526, and $\sim 11$ pc and $\sigma_{\rm rms} \approx 0.3$ K for LMC). We find that GMCs in NGC4526 tend to have higher velocity dispersions than equal size clouds in the LMC (Figure 13). With respect to the Milky Way disk, clouds in NGC4526 lie above the Milky Way disk size-linewidth relation (S87) by a factor of $\sim 3$ (Figure 13) and above the Milky Way size-luminosity relation by a factor of $\sim 5$ (Figure 5), which implies that clouds in NGC4526 are brighter and more turbulent than similarly-sized clouds in the Milky Way. In contrast, NGC4526 clouds are less turbulent than the Galactic center clouds by a factor of $\sim 0.4$ dex (Figure 13). This genuine variation of GMC properties may be influenced by different environment between galaxies \citep{hughes13} and is discussed in $\S 5.4$.

\subsection{Variations of GMC Surface Density}

\begin{figure}
%\figurenum{text}
\epsscale{1.2}
\plotone{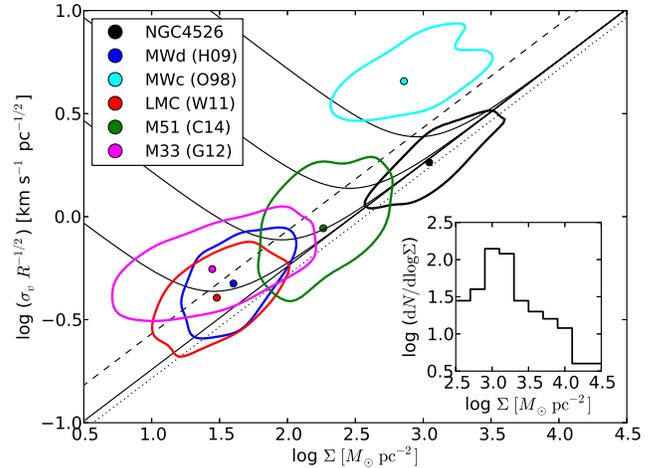}
\label{fig:14}
%\plottwo{epsfile}{epsfile}
\caption{Correlation between $\sigma_v R^{-1/2}$ and surface density for extragalactic GMC populations (color coded as in Figure 13). The contours enclose 68\% of the distribution of data points of a given galaxy. The solid `V`-curves are the pressure equilibrium condition of an isothermal sphere for various external pressures \citep[$P_{\rm ext}/k = 10^7, 10^6, 10^5, 10^4$, and 0 K cm$^{-3}$, respectively, from top to bottom;][]{field11}. The dashed line is the locus of critical surface density for a Bonnor-Ebert sphere. The dotted line is the gravitational equilibrium of a constant density sphere. This plot shows that the surface density of GMCs is not constant as previously believed. {\it Inset:} the surface density function of the clouds in NGC4526.}
\end{figure}

If the standard size-linewidth relation ($\sigma_v \propto R^{1/2}$) is valid for GMCs, then as a consequence the mass of gravitationally bound clouds is $M_{\rm vir} \propto R^2$ (c.f. equation 4), and hence the mass surface density $\Sigma = M_{\rm vir} / \pi R^2 = {\rm constant}$. However, \citet{heyer09}, who revisited the GMCs of S87 using more sensitive and better sampled data, found that the surface density is actually not constant, and the coefficient of the size-linewidth relation ($C_0 = \sigma_v R^{-1/2}$) correlates with the surface density as $C_0 \propto \Sigma^{1/2}$. This relation is expected from gravitational equilibrium (equation 4), and does not depend on whether the clouds follow the size-linewidth relation or not. The same relation ($C_0 \propto \Sigma^{1/2}$) also holds for pressure equilibrium, but with $\Sigma = \Sigma_{\rm c} \propto P_{\rm ext}^{1/2}$ (Field et al. 2011), where $\Sigma_{\rm c}$ is the critical surface density of a Bonnor-Ebert sphere \citep{bonnor56,ebert55}. The difference between the two is that pressure equilibrium has a higher normalization than the gravitational equilibrium. The \citet{heyer09} data favor pressure equilibrium rather than gravitational equilibrium \citep{field11}.

In Figure 14, we compile extragalactic GMC data. The contours enclose 68\% of the distribution of data points of each galaxy. Our compilation of extragalactic GMC data shows that the surface density is not constant, but varies from $\sim 10$ to $3000 \ M_{\odot} \ {\rm pc}^{-2}$. GMCs in the Milky Way disk, LMC, and M33 have lower surface densities than GMCs in M51 and NGC4526. The median surface density of the NGC4526 clouds is $\Sigma_{\rm med} \approx 1.2 \times 10^3 \ M_{\odot} \ {\rm pc}^{-2}$, which is $\sim 7$ times greater than in the Milky Way disk clouds ($170 \ M_{\odot} \ {\rm pc}^{-2}$; S87). However, NGC4526 clouds have similar surface density as the Galactic center clouds. This high surface density may be a common feature for clouds in the galaxy bulge.

In Figure 14, we also see that there is a correlation between $\sigma_v R^{-1/2}$ and surface density, as expected from gravitational (and pressure) equilibrium. The Milky Way disk, LMC, M33, M51, and NGC4526 clouds roughly follow $\sigma_v R^{-1/2} \propto \Sigma^{1/2}$, but the Milky Way, LMC, and M33 clouds have higher normalizations, i.e. they lie above the gravitational equilibrium relation (dotted line). As \citet{field11} suggested, Milky Way disk clouds are likely to be in pressure-equilibrium (dashed line) rather than gravitational equilibrium, and hence have a higher normalization factor. With the exception of Galactic center clouds, it is interesting to note the trend that GMCs with lower surface densities tend to be in pressure equilibrium. The Galactic center clouds are unique because they are pressure-bound clouds with $M_{\rm vir} \sim 10 \ M_{\rm lum}$ and they reside in the high external pressure environment \citep{oka98,miyazaki00}.

Based on this finding (Figure 14), we argue against the current mainstream view regarding the constancy of cloud surface density. Indeed, theoretical studies \citep[e.g.][]{kegel89,ballesteros02} found that limited observational sensitivities can give biased results, so that previous measurements of the surface density of GMCs appear constant. Moreover, size, velocity dispersion, and surface density are correlated with each other as expected from gravitational equilibrium (or pressure equilibrium as in the Milky Way, LMC, and M33).

Here, we propose a modified version of the Larson's relations. (1) The clouds are in either gravitational or pressure equilibrium; the relative contribution of gravity and external pressure to cloud stability need further study on a cloud-by-cloud basis. (2) The coefficient of the size-linewidth relation depends on the cloud surface density as $C_0 \propto \Sigma^{1/2}$. This relation also holds true in the pressure equilibrium case, where surface density depends on the external pressure of the ambient medium \citep[$\Sigma \propto P_{\rm ext}^{1/2}$;][]{field11}. (3) The cloud surface densities are not all the same, but may depend on the environment, such as the external pressure \citep{elmegreen93}, interstellar radiation field strength \citep{mckee89}, and interstellar gas flow and turbulence \citep{vazquez07}, which need further investigation. These environmental dependencies could explain variations of GMC properties across different galaxies \citep{hughes10,eros05}.

\subsection{Environmental Effects}

GMCs in NGC4526 are denser and more turbulent than those in the Galactic disk, but have similar surface density and less turbulent than those in the Galactic center (Figure 13 and 14). These differences may be caused by different environment, such as the interstellar radiation field strength (ISRF) and the external ambient pressure ($P_{\rm ext}$), between NGC4526 and the Milky Way. Here, we infer the ISRF and $P_{\rm ext}$ based on the global properties of the galaxy and discuss their possible roles to explain the differences between GMCs in NGC4526 and the Milky Way.

\subsubsection{Comparison with GMCs in the Milky Way disk}

\citet{ciesla14} have used the Herschel photometric data to derive the dust spectral energy distribution of 322 nearby galaxies, including NGC4526. They fit the data with the dust emission model of \citet{draine07}. In this model, a large fraction of dust is located in the diffuse interstellar medium, exposed to a single ISRF with intensity $U = U_{\rm min}$. We define $U$ as the intensity normalized to the Milky Way value, i.e. ${\rm ISRF} = U \times {\rm ISRF_{\rm MW}}$, where ${\rm ISRF_{\rm MW}}$ is the ISRF of the Milky Way \citep{mathis83}. In addition, there is a small fraction ($\gamma$) of dust located in regions where the ISRF is more intense (e.g. photodissociation regions), with ISRF ranging from $U_{\rm min}$ to $U_{\rm max}$ and described by a power law $U^{-\alpha}$. \citet{draine07} found $U_{\rm max} =10^6$ and $\alpha = 2$ are the best fit to the SINGS sample \citep{kennicutt03}. The free parameters of the model are then reduced to $U_{\rm min}$ and $\gamma$.\footnote{There is a third parameter of the model, namely the fraction of dust mass contributed by Polycyclic Aromatic Hydrocarbons (PAHs), but we do not need it to calculate the mean ISRF.} \citet{ciesla14} found the best fit parameters of $U_{\rm min} = 3.92 \pm 0.32$ and $\gamma = (0.19 \pm 0.09)\%$ for NGC4526.

Then, we can calculate the mean ISRF ($\bar U$) in NGC4526, weighted by the dust mass, by using equation (17) of \citet{draine07}:
\begin{equation}
\bar{U} = (1-\gamma) \ U_{\rm min} + \frac{\gamma \ {\rm ln}(U_{\rm max}/U_{\rm min})}{U_{\rm min}^{-1} - U_{\rm max}^{-1}} \approx 4.0.
\end{equation}
Thus, the ISRF in NGC4526 is $\sim 4$ times higher than in the Milky Way.

This higher ISRF means higher photoionization rate to the molecular gas, so that the CO emission emerges from a deeper layer within the cold gas, i.e. at higher extinction ($A_V$) than in the typical Milky Way disk. This higher extinction translates into higher gas surface density via \citep{mckee89}
\begin{equation}
\Sigma_{{\rm H}_2} = 22.3 \ \frac{A_V}{\delta_{\rm gr}} \ M_{\odot} \ {\rm pc}^{-2},
\end{equation}
where $\delta_{\rm gr}$ is the ratio of the extinction per hydrogen nucleus in the cloud to the standard value given by \citet{spitzer78}. Generally, $\delta_{\rm gr} = 1$ in the Milky Way. We do not know the value of $\delta_{\rm gr}$ in NGC4526, but we can assume that $\delta_{\rm gr}$ is proportional to the metallicity as $\delta_{\rm gr} \sim Z/Z_{\odot}$ \citep{bolatto08}. By using equation (23) and log$(Z/Z_{\odot}) \approx 0.2$ \citep{davis13c}, we can estimate the value of $A_V$ that is required to reproduce the observed surface density of GMCs in NGC4526. Cloud's surface density in NGC4526 is $\sim 7$ times higher than those values in the Milky Way disk (Figure 14), so that $A_V$ in NGC4526 is estimated to be 4.4 times higher than that in the Milky Way disk.

Furthermore, \citet{mckee89} predicts that the velocity dispersion of GMCs is proportional to the square-root of $A_V$ and the cloud size as $\sigma_v \propto (A_V/\delta_{\rm gr})^{1/2} \ R^{1/2}$. This relation arises naturally from the gravitational equilibrium state and by using $A_V$ as a proxy of surface density as in equation (23). By using the estimated value of $A_V$ above, the theory predicts that the velocity dispersion in NGC4526 clouds is about 2.1 times higher than the velocity dispersion of equal-size clouds in the Milky Way disk. This prediction is in agreement with our measurements, which show the velocity dispersion of equal-size clouds in NGC4526 is higher than those in the Milky Way disk by a factor of $\sim 3$ (Figure 5 and 13). Thus, we speculate that the surface density and velocity dispersion in NGC4526 clouds are higher because those clouds have higher extinction than clouds in the Milky Way disk.

\subsubsection{Comparison with GMCs in the Galactic center}

Clouds in NGC4526 have similar surface density and a smaller velocity dispersion than equal size clouds in the Galactic center \citep{oka98} by a factor of $\sim 0.4$ dex (Figure 13 and 14). This may be due to the fact that \citet{oka98} clouds and \citet{miyazaki00} clumps are in pressure equilibrium, rather than gravitational equilibrium as in NGC4526 clouds. From the equation (20), by neglecting the magnetic pressure term, this means the internal pressure of pressure-bound clouds needs to balance against gravity and external pressure, while gravitationally bound clouds needs to balance against gravity only. Therefore, for a given cloud mass and radius, the velocity dispersion of pressure-bound clouds is higher than the velocity dispersion of gravitationally bound clouds in order to maintain a dynamical equilibrium state.

We do the following calculations to support our argument. By neglecting the magnetic field, gravitationally bound clouds have $\sigma_{v,{\rm vir}}^2 = P_G \ \rho^{-1}$, while pressure-bound clouds have $\sigma_{v,{\rm pres}}^2 = (P_G + P_{\rm ext}) \ \rho^{-1}$, where $\rho \propto \Sigma \ R^{-1}$ and $P_G \propto \Sigma^2$ (BM92). For equal-size clouds with similar density, the ratio between the two is
\begin{equation}
\frac{\sigma_{v,{\rm pres}}}{\sigma_{v,{\rm vir}}} = \left( 1 + \frac{P_{\rm ext}}{P_G}\right)^{1/2}.
\end{equation}

For spherical clouds, like those in NGC4526, with mass $\sim 10^6 M_{\odot}$ and radius $\sim 20$ pc, $P_G$ is $\sim 8 \times 10^6$ K cm$^{-3}$. The external pressure $P_{\rm ext}$ in the Galactic center is rather uncertain. For $P_{\rm ext}/k$ between $\sim 5 \times 10^6$ \citep{spergel92} and $\sim 1 \times 10^8$ \citep{miyazaki00}, we get the Galactic center clouds are expected to have higher velocity dispersion by a factor of 1.3 to 3.7 (c.f. equation 24). This range is also in agreement with our measurement ($\sim 0.4$ dex $\approx 2.5$; Figure 13).

If this is true, then why are the clouds in NGC4526 bulge in gravitational equilibrium but clouds in the Galactic center pressure bound? To get insight into this question, we estimate the global ambient hydrostatic pressure $P_h$ in NGC4526 as a proxy of the external pressure \citep{elmegreen89}:
\begin{equation}
P_{\rm h} = \frac{\pi G}{2} \Sigma_g \left( \Sigma_g + \frac{\sigma_g}{\sigma_*} \ \Sigma_* \right),
\end{equation}
where $\Sigma_g \equiv \Sigma_{\rm HI}$ is the ambient gas surface density, $\Sigma_*$ is the stellar surface density, $\sigma_g$ is the ambient gas velocity dispersion, and $\sigma_*$ is the stellar velocity dispersion.

We use ATLAS-3D \citep{cappellari11} results to get the stellar properties of NGC4526: $\sigma_* \approx 233.3$ km/s (at the central 1 kpc), $L \approx 3.13 \times 10^{10} L_{\odot,{\rm r}}$ \citep{cappellari13a}, stellar $M/L_{\rm r} \approx 5.6 \ M_{\odot}/L_{\odot,{\rm r}}$ \citep{cappellari13b}, and $R_e \approx 74.1" \approx 5.9$ kpc \citep{krajnovic13}. The quoted luminosity value is global, so the enclosed luminosity at $R_e$ is simply half the quoted value, i.e. $L(R_e) \approx 1.57 \times 10^{10} \ L_{\odot}$. Thus, the stellar surface density can be estimated as $\Sigma_* \approx M(R_e)/\pi R_e^2 \approx 804 \  M_{\odot} \ {\rm pc}^{-2}$.

HI is undetected in NGC4526 with upper limit of $M_{\rm HI} < 1.9 \times 10^7 \ M_{\odot}$ \citep{lucero13}. The linear resolution of their observations is $5.4 \times 4.2$ kpc$^2$. This gives an upper limit of the gas surface density as $\Sigma_g < 0.27 \ M_{\odot} \ {\rm pc}^{-2}$. We do not know the value of $\sigma_g$, so we assume $\sigma_g = 10$ km/s \citep{blitz06}.

Taken all together, we estimate the external pressure to be $P_h / k < 1.5 \times 10^6$ K cm$^{-3}$. Thus, unlike the Galactic center clouds, where $P_{\rm ext} \sim P_G$, NGC4526 clouds have $P_{\rm ext} < P_G$. This may cause the Galactic center clouds are pressure bound, while clouds in NGC4526 are gravitationally bound (since the external pressure is small with respect to the gravity). This small external pressure is presumably due to lack of HI in NGC4526, which may be caused by ram pressure or hot gas evaporation as the galaxy resides in the Virgo cluster.

\section{Summary}

We identify 241 GMCs based on $^{12}$CO($2-1$) observation at $\approx 20$ pc resolution in the galaxy NGC4526 using the CPROPS program (RL06), where 103 of them are spatially resolved. As a population, the clouds are in gravitational equilibrium. A log-normal fit to the population yields a mean virial parameter $\alpha \approx 0.99$ with a standard deviation of $\sim 0.14$ dex.

The cloud mass distribution follow d$N/$d$M \propto M^{-2.39 \pm 0.03}$, steeper than in the inner Milky Way but comparable to what others find in several other galaxies. Since the exponent is less than $-2$, the total molecular mass is dominated by the contribution of low mass clouds. The data favor a truncated-distribution with truncation mass of $4.12 \times 10^6 M_{\odot}$.

In general, clouds in NGC4526 are more luminous and more turbulent than equal-size clouds in the Milky Way disk by a factor of $\sim 5$ and $\sim 3$, respectively. Moreover, the surface density of GMCs in NGC4526 is $\sim 7$ times higher than those in the Milky Way disk. These differences may be due to higher ISRF and cloud extinction ($A_V$) in NGC4526, so that the CO emission emerges from a deeper layer in the cold gas, and hence, a higher gas density region.

On the other hand, NGC4526 clouds are less turbulent than the Galactic center clouds. This may be caused by different equilibrium state of GMCs: Galactic center clouds are pressure-bound, while clouds in NGC4526 are gravitationally bound. The velocity dispersion of the pressure-bound clouds needs to balance both gravity and the external pressure, while gravitationally bound clouds just need to balance gravity only. Indeed, our estimation shows that the external pressure in NGC4526 is smaller than the gravity, so that the external pressure is less important in the dynamical state of NGC4526 clouds. This situation is different in the Galactic center, where the external pressure is comparable or higher than the cloud self-gravity.

We find no size-linewidth correlation in NGC4526 in contrast to what is expected from Larson's relation. This finding is robust against the choice of the input parameters of the CPROPS program or different measurement methods (Appendix D). This implies that the surface density of GMCs is not constant, but follows the relation $\sigma_{v} R^{-1/2} \propto \Sigma^{1/2}$ as expected from gravitational equilibrium.

In the kinematic analysis, we find that the velocity gradient of individual clouds are just a consequence of galactic rotation. In this case, if the clouds are rotating, then the rotation follows the galactic shear given by the Oort $A$ constant at the location of the cloud. We calculate $\Omega_{\rm shear}$ and find that 92\% of resolved clouds have a turbulent energy exceeding the rotational energy, and 96\% of resolved clouds have a gravitational energy exceeding the rotational energy. This means rotational energy is a minor contribution to the clouds' dynamical stability.

Even with the inclusion of rotational energy, the cloud population is still in gravitational equilibrium. The distribution of the virial parameter can be approximated by a log-normal distribution with a tail at the high-end. The mean of the distribution is $\alpha \approx 1.26$ with a standard deviation of $\sim 0.15$ dex. There are only 4 clouds with $\alpha \gtrsim 3.5$. These clouds are the innermost clouds and undergo extreme galactic shear. These unbound clouds should be ripped apart in a timescale of less than $\sim 1$ Myr.

\acknowledgments

We thank Christopher McKee, Richard Plambeck, Tom Hartquist, Eve Ostriker, and Adam Leroy for valuable discussions. We acknowledge a constructive report from the anonymous referee, which greatly improves the paper. This work was partially supported by a grant from the National Science Foundation. ER is supported by a Discovery Grant from NSERC of Canada. MB acknowledge support from the ATLAS-3D project. MC acknowledge support from a Royal Society University Research Fellowship. Support for CARMA construction was derived from the states of California, Illinois, and Maryland, the James S. McDonnell Foundation, the Gordon and Betty Moore Foundation, the Kenneth T. and Eileen L. Norris Foundation, the University of Chicago, the Associates of the California Institute of Technology, and the National Science Foundation. Ongoing CARMA development and operations are supported by the National Science Foundation under a cooperative agreement, and by the CARMA partner universities. This research has made use of the NASA/IPAC Extragalactic Database (NED) which is operated by the Jet Propulsion Laboratory, California Institute of Technology, under contract with the National Aeronautics and Space Administration.

\appendix

\section{A. Cloud Identification Algorithm}

We use the modified CLUMPFIND algorithm \citep{williams94}, implemented in the CPROPS program \citep[hereafter RL06]{eros06} to identify GMCs in NGC4526. Below are descriptions of the code, together with our chosen values of input parameters.

First, the program identifies connected regions of significant emission as {\it islands}. An island is defined as CO emission that has at least one pixel higher than $3 \sigma_{\rm rms}$ (the threshold parameter of the program) and extends to all connected pixels with emission higher than $2 \sigma_{\rm rms}$ (the edge parameter of the program). An island consists of one or more clouds after the decomposition process. We set the minimum volume of islands to be 1 beamwidth$^2 \times 1$ velocity channel. We choose these values to include any possible small island in our data, since our resolution is somewhat comparable to the typical size of Milky Way GMCs.

The decomposition of each island begins by looking for local maxima. Local maxima are identified by looking for pixels that are greater than or equal to all neighbors in a 1 beamwidth$^2$ $\times$ 1 velocity channel volume. Our choice of these values is to separate an island into potentially smaller clouds.

For each local maximum, working from the local maximum that has the lowest emission to the highest, the algorithm identifies pixels associated {\it exclusively} with each local maximum by contouring the data cube in three dimensions. If the emission of a local maximum is less than $n \sigma_{\rm rms}$ above the {\it merge level} with neighboring local maxima, or there are fewer than $m \times$beamwidth pixels associated with the local maxima, then the local maximum is removed from consideration. The merge level is the contour level that encloses two neighboring local maxima. The purpose of this decimation process is to remove spurious peaks of noise, i.e. false clouds. Higher values of $n$ and $m$ gives a smaller probability of false clouds, at the cost of losing small genuine clouds. We adopt $m = 0.5$ to account for small clouds and determine the best value of $n$ from simulations (described in Appendix B).

Then, the program decides whether two neighboring clouds are a merged cloud or distinct clouds. The algorithm compares the values of emission moments for the separated and combined clouds. If the flux $F$ and moment $\sigma$ of an individual cloud differ by more than a fraction of the flux and the moment of the merged cloud ($\delta F/F$ and $\delta \sigma/\sigma$), then the local maxima are categorized as distinct. The values of $\delta F/F$ and $\delta \sigma/\sigma$ are chosen from the simulations (Appendix B).

At the end of the process, the program calculates the properties of the clouds (described in $\S 3.1$) and records them into a catalog (Table 1).

\section{B. Decomposition Parameters}

In order to choose the best decomposition parameters ($n, \delta F/F$, and $\delta \sigma/\sigma$) that are suitable for our data, we create two simulated clouds with a 3D-Gaussian shape in a single data cube, and add the typical noise of our observations. The dispersions of the Gaussian are $\sigma_{\rm x} = \sigma_{\rm y} = 1$ beam width and $\sigma_v = 1$ velocity channel, so that they are resolved by the antenna beam and spectral channels. We vary the separation (center-to-center) of those gaussians in units of $R$, which is defined as $R = 2\sigma_{\rm x}$. Two Gaussians are almost fully resolved if they are separated by a distance larger than $2R$. We also vary the peak S/N of the Gaussians from $3\sigma_{\rm rms}$ to $6\sigma_{\rm rms}$, to take into account any possible dependence of our simulations on S/N. We run 10 simulations for each choice of separation distance and peak S/N, so that the results are statistically robust.

\begin{figure}
%\figurenum{text}
\epsscale{0.6}
\plotone{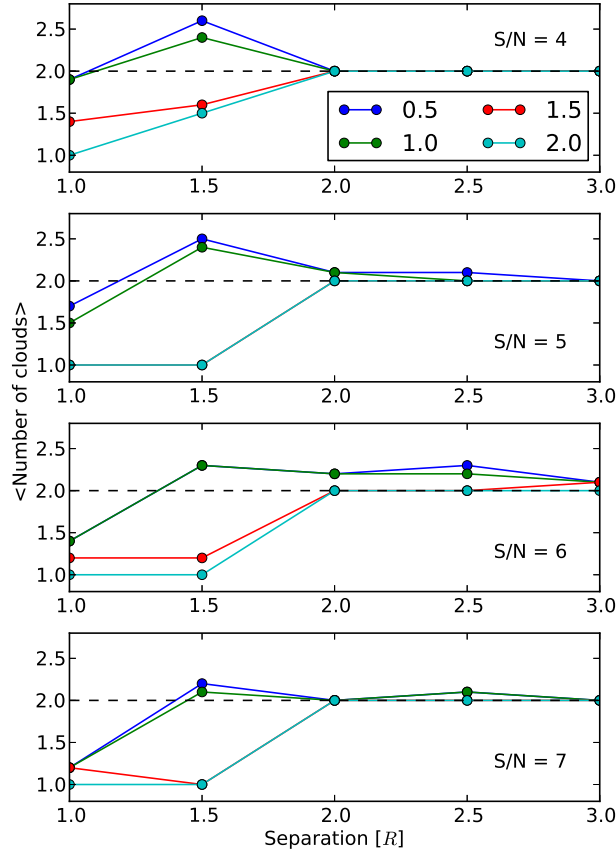}
\label{fig:15}
%\plottwo{epsfile}{epsfile}
\caption{Average number of recovered clouds as a function of the separation (center-to-center) between two clouds. The separation distance is in units of $R = 2\sigma_{\rm x}$, where $\sigma_{\rm x}$ is the dispersion of the Gaussian clouds. Different panels show different peak S/N of the simulated clouds. In each panel, different values of the decomposition parameters ($n$, $\delta F/F$, and $\delta \sigma/\sigma$) are shown in different colors. The correct number of clouds is indicated by the dashed lines. The program recovers approximately two clouds for separations of at least $2R$. Unity input parameters approach the correct number of clouds even for blended clouds.}
\end{figure}

We feed the simulated data cubes into the CPROPS program and vary the three input parameters that drive the decomposition of the clouds ($n$, $\delta F/F$, and $\delta \sigma/\sigma$) from 0 to 3 with an increment of 0.5. The program then identifies the number of clouds in a given data cube. For various decomposition parameters and peak S/N, we plot the average number of clouds identified by the program against the separation distance in Figure 15. The program successfully resolves two clouds for a separation distance larger than $2R$. However, for blended Gaussians (separated by a distance shorter than $2R$), the values of $n = \delta F/F = \delta \sigma/\sigma = 1$ best recover the correct number of clouds at all S/N. Therefore, we adopt these values as the decomposition parameters for our data. In Appendix D, we further show that the results of our studies are not sensitive to the choice of decomposition parameters.

\section{C. Probability Analysis of Real Detections}

In order to check the probability that the identified clouds are real, we do a probability analysis similar to that given in \citet{engargiola03}. If we have $n$-adjacent channels with the same brightness temperature $T_{\rm b} = k\sigma_{\rm rms}$, then the probability of this being a false detection is $P_n(k) = [0.5 \times {\rm erfc}(k/\sqrt{2})]^n$, where erfc is the Gaussian complementary error function. The probability of real detection is $P_{\rm real} = 1 - N_{\rm trial} P_n(k)$ for $N_{\rm trial}P_n(k) \ll 1$. Here, $N_{\rm trial} = N/n$, where $N \approx 2.37 \times 10^7$ is the number of pixels in our data cube. If the pixels are not independent due to beam convolution and spectral smoothing, then the inferred $P_{\rm real}$ is smaller than the true $P_{\rm real}$. We set the edge parameter of the CPROPS program to be $2\sigma_{\rm rms}$, so that all pixels in a cloud must have $T_b \geq 2\sigma_{\rm rms}$. Hence, the probability that a cloud occupying $n$-adjacent channels is a real detection is $P_{\rm real} > 1-N_{\rm trial} P_n(2\sigma_{\rm rms})$. In this case, $P_{\rm real} > 0.97$ for $n = 5$. The smallest identified cloud has total number of pixels $n = 13$. This suggests that it is unlikely that we detect false clouds, so we assume all identified GMCs are real structures.

\section{D. Checking Bias Against the Choice of Input Parameters}

\begin{figure}
%\figurenum{text}
\epsscale{1.2}
\plotone{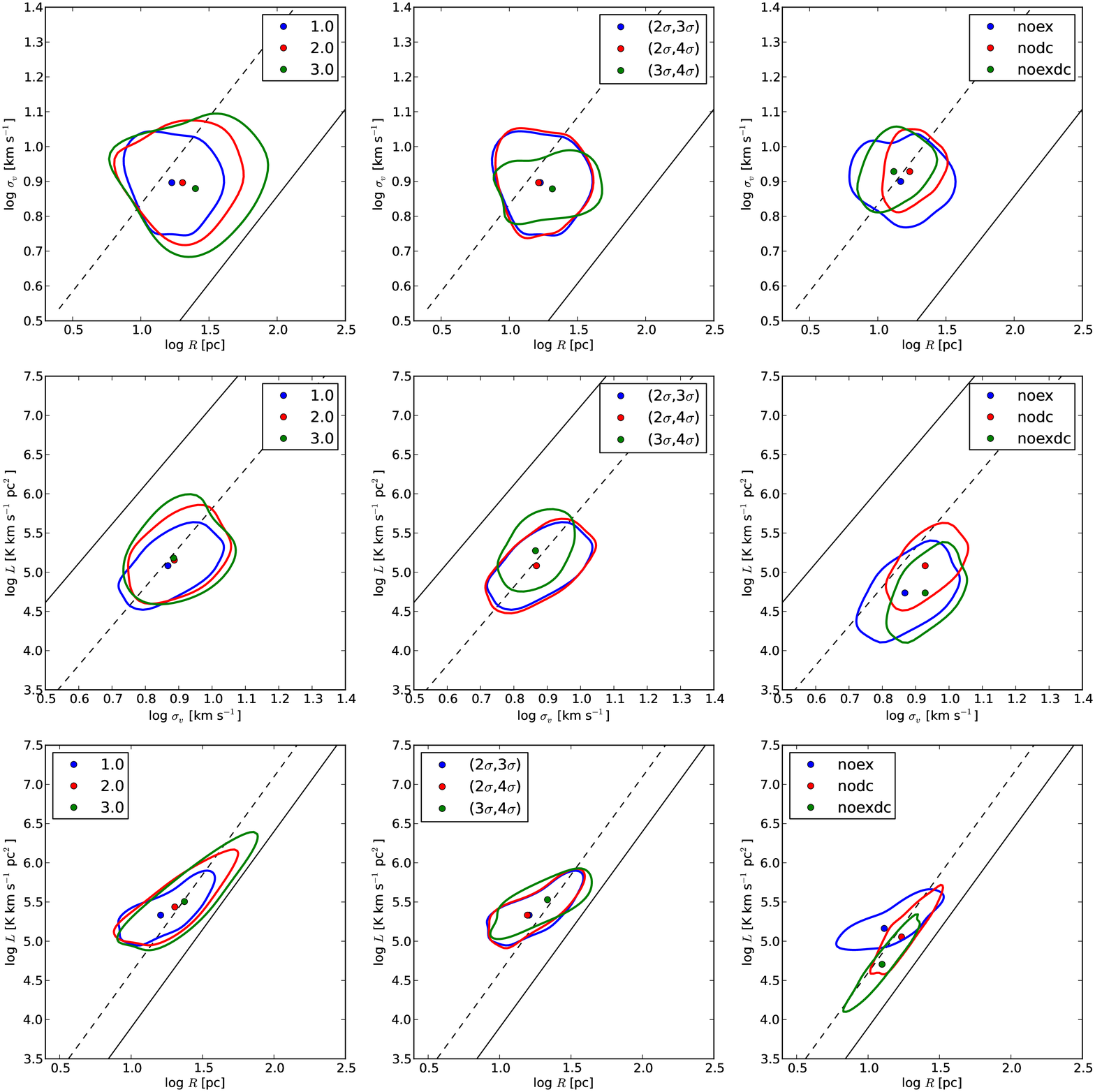}
\label{fig:16}
%\plottwo{epsfile}{epsfile}
\caption{Plots of GMC properties (radius, velocity dispersion, and luminosity) for various decomposition parameters (left), islands parameters (middle), and methods of measurement (right). The black lines are Larson's relations and the dashed lines are Larson's relations with higher or lower normalization factors. There is no obvious bias due to the choice of input parameters of the program.}
\end{figure}

We evaluate how our results are affected by the choice of input parameters of the CPROPS program (as described in Appendix B) and the methods to measure the cloud properties (as described in $\S 3$). We rerun the program and recalculate the cloud properties (radius, velocity dispersion, and luminosity) using various decomposition parameters ($n,\delta F/F$, and $\delta \sigma/\sigma$), edge and threshold parameters for islands, and methods of measurement (by excluding deconvolution, excluding extrapolation, and excluding both deconvolution and extrapolation).

\begin{table}
\caption{Spearman Correlation Coefficients for Various Input Parameters and Measurement Methods}
\centering
\begin{tabular}{ c c c c }
\tableline
Parameters & $R$ vs. $\sigma_v$ & $\sigma_v$ vs. $L$ & $R$ vs. $L$ \\[0.5ex]
\tableline
$n = \delta F/F = \delta \sigma/\sigma$ = 1 & $-0.14$ & 0.51 & 0.67 \\[0.5ex]
$n = \delta F/F = \delta \sigma/\sigma$ = 2 & \phantom{1}$0.04$ & 0.45 & 0.81 \\[0.5ex]
$n = \delta F/F = \delta \sigma/\sigma$ = 3 & \phantom{1}$0.09$ & 0.37 & 0.87 \\[0.5ex]
\tableline
(edge,threshold) $= (2\sigma,3\sigma)$ & $-0.14$ & 0.51 & 0.67 \\[0.5ex]
(edge,threshold) $= (2\sigma,4\sigma)$ & $-0.09$ & 0.53 & 0.69 \\[0.5ex]
(edge,threshold) $= (3\sigma,4\sigma)$ & \phantom{1}$0.06$ & 0.40 & 0.68 \\[0.5ex]
\tableline
no extrapolation & $-0.12$ & 0.48 & 0.68 \\[0.5ex]
no deconvolution & \phantom{1}$0.21$ & 0.51 & 0.83 \\[0.5ex]
no extrapolation \& no deconvolution & \phantom{1}$0.28$ & 0.48 & 0.94 \\[0.5ex]
\tableline
\end{tabular}
\end{table}

\begin{table}
\caption{Gaussian Fit Coefficients of the Distributions in Figure 16}
\centering
\begin{tabular}{ c | c c c | c c c }
\tableline
\multirow{2}{*}{Parameters} & \multicolumn{3}{ c }{Gaussian center} & \multicolumn{3}{ | c }{Gaussian dispersion} \\[0.5ex]
 & $R$ vs. $\sigma_v$ & $\sigma_v$ vs. $L$ & $R$ vs. $L$ & $R$ vs. $\sigma_v$ & $\sigma_v$ vs. $L$ & $R$ vs. $L$ \\[0.5ex]
\tableline
$n = \delta F/F = \delta \sigma/\sigma$ = 1 & (34.39, 14.14) & (16.98, \phantom{1}8.04) & (31.05, 13.04) & (24.11, \phantom{1}8.47) & (\phantom{1}6.86, \phantom{1}8.67) & (16.72, \phantom{1}9.27) \\[0.5ex]
$n = \delta F/F = \delta \sigma/\sigma$ = 2 & (11.97, 14.88) & (16.12, \phantom{1}1.40) & (14.85, \phantom{1}3.86) & (11.62, 10.62) & (\phantom{1}8.73, \phantom{1}5.97) & (\phantom{1}7.02, \phantom{1}5.91) \\[0.5ex]
$n = \delta F/F = \delta \sigma/\sigma$ = 3 & (12.50, 13.95) & (13.93, \phantom{1}1.25) & (16.61, \phantom{1}4.74) & (14.46, 12.73) & (10.29, \phantom{1}7.94) & (\phantom{1}7.75, \phantom{1}7.26) \\[0.5ex]
(edge,threshold) $= (2\sigma,3\sigma)$ & (34.39, 14.14) & (16.98, \phantom{1}8.04) & (31.05, 13.04) & (24.11, \phantom{1}8.47) & (\phantom{1}6.86, \phantom{1}8.67) & (16.72, \phantom{1}9.27) \\[0.5ex]
(edge,threshold) $= (2\sigma,4\sigma)$ & (33.49, 14.42) & (16.83, \phantom{1}7.74) & (30.00, 13.09) & (24.95, \phantom{1}8.94) & (\phantom{1}7.17, \phantom{1}9.11) & (15.70, \phantom{1}9.51) \\[0.5ex]
(edge,threshold) $= (3\sigma,4\sigma)$ & (51.39, 27.88) & (38.92, 11.43) & (52.06, 31.64) & (32.53, 14.59) & (\phantom{1}9.08, 21.79) & (25.97, 16.04) \\[0.5ex]
no extrapolation & (28.00, 34.30) & (17.00, \phantom{1}5.54) & (25.50, 12.18) & (35.47, 17.38) & (\phantom{1}6.86, \phantom{1}7.71) & (22.53, 12.72) \\[0.5ex]
no deconvolution & (32.10, 15.19) & (14.96, \phantom{1}7.99) & (30.12, \phantom{1}7.65) & (14.26, \phantom{1}6.91) & (\phantom{1}6.54, \phantom{1}8.68) & (\phantom{1}9.50, \phantom{1}6.04) \\[0.5ex]
no extrapolation \& no deconvolution & (24.78, 15.22) & (14.99, \phantom{1}5.50) & (23.20, \phantom{1}5.70) & (14.90, \phantom{1}6.88) & (\phantom{1}6.55, \phantom{1}7.71) & (\phantom{1}8.77, \phantom{1}4.66) \\[0.5ex]
\tableline
\end{tabular}
\end{table}

\begin{table}
\caption{Two-Dimensional Kolmogorov-Smirnov Test}
\centering
\begin{tabular}{ c c c c | c c c c | c c c c }
\tableline
\multicolumn{12}{c}{$R$ vs. $\sigma_v$} \\[0.5ex]
\tableline
Parameters & 1 & 2 & 3 & Parameters & ($2\sigma,3\sigma$) & ($2\sigma,4\sigma$) & ($3\sigma,4\sigma$) & Parameters & noex & nodc & noexdc \\[0.5ex]
1 & 1.00 & - & - & ($2\sigma,3\sigma$) & 1.00 & - & - & noex & 1.00 & - & - \\[0.5ex]
2 & 0.11 & 1.00 & - & ($2\sigma,4\sigma$) & 1.00 & 1.00 & - & nodc & 0.00 & 1.00 & - \\[0.5ex]
3 & 0.01 & 0.72 & 1.00 & ($3\sigma,4\sigma$) & 0.05 & 0.04 & 1.00 & noexdc & 0.08 & 0.00 & 1.00 \\[0.5ex]
\tableline
\multicolumn{12}{c}{$\sigma_v$ vs. $L$} \\[0.5ex]
\tableline
Parameters & 1 & 2 & 3 & Parameters & ($2\sigma,3\sigma$) & ($2\sigma,4\sigma$) & ($3\sigma,4\sigma$) & Parameters & noex & nodc & noexdc \\[0.5ex]
1 & 1.00 & - & - & ($2\sigma,3\sigma$) & 1.00 & - & - & noex & 1.00 & - & - \\[0.5ex]
2 & 0.13 & 1.00 & - & ($2\sigma,4\sigma$) & 1.00 & 1.00 & - & nodc & 0.00 & 1.00 & - \\[0.5ex]
3 & 0.15 & 0.74 & 1.00 & ($3\sigma,4\sigma$) & \phantom{1}0.005 & \phantom{1}0.004 & 1.00 & noexdc & 0.00 & 0.00 & 1.00 \\[0.5ex]
\tableline
\multicolumn{12}{c}{$R$ vs. $L$} \\[0.5ex]
\tableline
Parameters & 1 & 2 & 3 & Parameters & ($2\sigma,3\sigma$) & ($2\sigma,4\sigma$) & ($3\sigma,4\sigma$) & Parameters & noex & nodc & noexdc \\[0.5ex]
1 & 1.00 & - & - & ($2\sigma,3\sigma$) & 1.00 & - & - & noex & 1.00 & - & - \\[0.5ex]
2 & 0.07 & 1.00 & - & ($2\sigma,4\sigma$) & 1.00 & 1.00 & - & nodc & 0.00 & 1.00 & - \\[0.5ex]
3 & 0.01 & 0.74 & 1.00 & ($3\sigma,4\sigma$) & 0.02 & 0.02 & 1.00 & noexdc & 0.00 & 0.00 & 1.00 \\[0.5ex]
\tableline
\end{tabular}
\end{table}

The decomposition parameters are varied from 1 to 3 with unity increment (first column of Figure 16), the edge parameter varies from $2\sigma_{rms}$ to $3\sigma_{\rm rms}$, and the threshold parameter varies from $3\sigma_{\rm rms}$ to $4\sigma_{\rm rms}$ (second column of Figure 16). Different methods of measurements are given in the third column of Figure 16. Each row in Figure 16 shows the plots between various cloud properties, i.e. size vs. linewidth in the first row, linewidth vs. luminosity in the second row, and size vs. luminosity in the third row. Larson's relations for the Milky Way are shown as the solid lines. For comparison, the dashed lines are Larson's relations with different normalization factors: 3, 0.05, and 5 for $R$ vs. $\sigma_v$, $\sigma_v$ vs. $L$, and $R$ vs. $L$, respectively. To check how strong the correlation between various cloud properties are, we tabulate the Spearman correlation coefficients $r_{\rm sp}$ in Table 3.

For each plot, we build an estimate of the probability density function (PDF) based on the data scatter in two-dimension, using the kernel-density-estimate method in Scipy. The contours in Figure 16 enclose to the 68\% confidence level of the PDFs. We also fit the PDFs with a 2D-Gaussian and show the Gaussian centers as filled circles in Figure 16. The center and dispersion of the Gaussians are tabulated in Table 4. For clarity, we do not show the data points. In addition, we check whether the PDFs in a given panel of Figure 16 are sampled from the same parent distribution or not. This is tested with the two-dimensional Kolmogorov-Smirnov (KS) probability value (Table 5). The higher the probability the more likely the PDFs are drawn from the same parent distribution.

For the first column in Figure 16, we see that higher decomposition parameters (e.g. $n = \delta F/F = \delta \sigma/\sigma = 3$, green contours) yield larger scatters than lower decomposition parameters (red and blue contours). The number of clouds for higher decomposition parameters is less than for lower decomposition parameters, so this larger scatter is not due to a larger number of data points. The larger scatter is probably due to the tendency of the program to combine small, neighboring clouds into bigger, merged clouds in the molecular ring island, while leaving the outer small islands as small clouds (Figure 1). This tendency can be seen in Figure 16, as the PDFs of higher decomposition parameters extend to larger radii and higher luminosities. The Gaussian centers of higher decomposition parameter also have larger radii and higher luminosities than those of lower decomposition parameters. In any case, none of those distributions shows a size-linewidth relation ($-0.14 \leq r_{\rm sp} \leq 0.09$), and they have different normalization factors than Larson's relations.

For the second column in Figure 16, the parameters (edge, threshold) = ($2\sigma_{\rm rms},3\sigma_{\rm rms}$) almost have the same PDF as (edge, threshold) = ($2\sigma_{\rm rms},4\sigma_{\rm rms}$). The KS-test yields a probability value of $\sim 1$. This is because we only lose a few clouds with peak S/N $< 4\sigma_{\rm rms}$. However, for parameters (edge, threshold) = ($3\sigma_{\rm rms},4\sigma_{\rm rms}$), we lose many clouds because the islands only extend to connected pixels with $T_b > 3\sigma_{\rm rms}$ (i.e. the islands get smaller), and hence the spread of the PDF decreases due to a smaller number of data points. None of the distributions shows a size-linewidth relation, and they have different normalization factors than Larson's relations.

For the third column in Figure 16, there is no distribution that yields the size-linewidth relation found in the Milky Way. In the linewidth vs luminosity plot, the non-deconvolved distributions (red and green contours) yield larger velocity dispersions, as expected from equation (2) by excluding the $\delta v^2/2\pi$ term, and the extrapolated distribution (red contour) yields higher luminosities. All distributions in the linewidth vs. luminosity plot lie below Larson's relation. In the size vs luminosity plot, the non-deconvolved distribution (red contour) yields a larger size, as expected from equation (1) by excluding the $\sigma_{\rm beam}^2$ terms. All distributions in the size vs. luminosity plot lie above Larson's relation.

Overall, we conclude that there is no significant effect on our general results due to the choice of input parameters and measurement methods. In particular, the absence of a size-linewidth relation and the different normalization factors of Larson's relations are still remain.

\end{document}